\RequirePackage[british]{babel}
\documentclass[reqno,a4paper,12pt,final]{amsart}
\usepackage{a4wide}
\usepackage[utf8x]{inputenc}
\usepackage{amsmath,amssymb,amstext,amsthm,amscd,mathrsfs,eucal}
\usepackage{graphicx,color}
\usepackage{array}
\usepackage{cite}
\usepackage{hyperref}
\hypersetup{%
  pdftitle   = {Metric Lie 3-algebras in Bagger-Lambert theory},
  pdfkeywords = {lorentzian, M2, M Theory, Bagger-Lambert, Lie, Filippov, 3-algebra},
  pdfauthor  = {José Figueroa-O'Farrill, Paul de Medeiros, Elena Méndez-Escobar},
  pdfcreator = {\LaTeX\ with package \flqq hyperref\frqq}
}
\PrerenderUnicode{éÉ}
\newcommand{\half}{\tfrac12}
\newcommand{\fg}{\mathfrak{g}}
\newcommand{\fa}{\mathfrak{a}}
\newcommand{\fb}{\mathfrak{b}}

\newcommand{\fD}{\mathfrak{D}}

\newcommand{\fgl}{\mathfrak{gl}}
\newcommand{\fh}{\mathfrak{h}}

\newcommand{\fk}{\mathfrak{k}}
\newcommand{\fl}{\mathfrak{l}}
\newcommand{\fm}{\mathfrak{m}}

\newcommand{\fr}{\mathfrak{r}}
\newcommand{\fs}{\mathfrak{s}}
\newcommand{\ft}{\mathfrak{t}}
\newcommand{\fw}{\mathfrak{w}}
\newcommand{\fso}{\mathfrak{so}}
\newcommand{\fosp}{\mathfrak{osp}}

\newcommand{\fu}{\mathfrak{u}}
\newcommand{\fzero}{\mathfrak{0}}
\newcommand{\SO}{\mathrm{SO}}

\newcommand{\GL}{\mathrm{GL}}

\newcommand{\RR}{\mathbb{R}}

\newcommand{\ZZ}{\mathbb{Z}}
\newcommand{\eL}{\mathscr{L}}

\newcommand{\be}{\boldsymbol{e}}

\newcommand{\Cbar}{\overline{C}}
\newcommand{\Ubar}{\overline{U}}
\newcommand{\Wbar}{\overline{W}}
\DeclareMathOperator{\Aut}{Aut}

\DeclareMathOperator{\End}{End}
\DeclareMathOperator{\Der}{Der}
\DeclareMathOperator{\Ad}{Ad}
\DeclareMathOperator{\ad}{ad}
\DeclareMathOperator{\Rad}{Rad}
\DeclareMathOperator{\id}{id}
\DeclareMathOperator{\im}{im}

\DeclareMathOperator*{\Cyc}{\mathscr{C}}
\theoremstyle{plain}
\newtheorem{lemma}{Lemma}
\newtheorem{proposition}[lemma]{Proposition}
\newtheorem{theorem}[lemma]{Theorem}
\newtheorem{corollary}[lemma]{Corollary}
\theoremstyle{definition}
\newtheorem{definition}[lemma]{Definition}
\newtheorem{remark}[lemma]{Remark}
\newcommand{\MUNCH}[1]{\relax}

\allowdisplaybreaks
\begin{document}
\title{Metric Lie 3-algebras in Bagger-Lambert theory}
\author{Paul de Medeiros, José Figueroa-O'Farrill and Elena Méndez-Escobar}
\address{Maxwell Institute and School of Mathematics, University of Edinburgh, UK}
\email{P.deMedeiros@ed.ac.uk, J.M.Figueroa@ed.ac.uk, E.Mendez@ed.ac.uk}
\date{\today}
\begin{abstract}
  We recast physical properties of the Bagger--Lambert theory, such as
  shift-symmetry and decoupling of ghosts, the absence of scale and
  parity invariance, in Lie 3-algebraic terms, thus motivating the study
  of metric Lie 3-algebras and their Lie algebras of derivations.  We
  prove a structure theorem for metric Lie 3-algebras in arbitrary
  signature showing that they can be constructed out of the simple and
  one-dimensional Lie 3-algebras iterating two constructions: orthogonal
  direct sum and a new construction called a double extension, by
  analogy with the similar construction for Lie algebras.  We classify
  metric Lie 3-algebras of signature $(2,p)$ and study their Lie
  algebras of derivations, including those which preserve the conformal
  class of the inner product.  We revisit the 3-algebraic criteria spelt
  out at the start of the paper and select those algebras with signature
  $(2,p)$ which satisfy them, as well as indicate the construction of
  more general metric Lie 3-algebras satisfying the ghost-decoupling
  criterion.
\end{abstract}
\maketitle
\tableofcontents

\section{Introduction}
\label{sec:introduction}

The foundational work of Bagger and Lambert \cite{BL1} and Gustavsson
\cite{GustavssonAlgM2}, has led to the construction in \cite{BL2} of a
superconformal field theory in three-dimensional Minkowski spacetime as
a model of multiple M2-branes in M-theory.  This theory realises the
maximal $\fosp(8|4)$ superconformal symmetry of the near-horizon
geometry of M2-branes in eleven-dimensional supergravity.  The lagrangian
and its equations of motion nicely encapsulate several other features
expected \cite{SchwarzChernSimons, BasuHarvey} in the long sought-after
low-energy description of multiple M2-branes.  These encouraging
properties have prompted a great deal of interest in the Bagger--Lambert
theory \cite{BL3, MukhiBL, BandresCS, BermanAMM, VanRaamsdonkBL,
  MorozovMM2, LambertTong, DistlerBL, GranNilssonBL, Ho3Lie,
  BergshoeffMM2, GP3Lie, GG3Lie, HoM5M2, GMRBL, BRGTV, HIM-M2toD2rev,
  MorozovBLG, MasahitoBL, HIM-M5MM2, Krishnan, JKKKP, LiWang,
  BanerjeeSenBL, LinBL, Lor3Lie, Gustavsson1loop, BLSNoGhost,
  ParkSochichiu, Passerini, GomisSCFT, MaldacenaBL, KlebanovBL,
  D2toD2, CecottiSenBL, MauriPetkouBL,BdRHR-BL}.

The Bagger--Lambert theory has a novel local gauge symmetry which is not
based on a Lie algebra, but rather on a Lie 3-algebra \cite{Filippov}.
The analogue of the Lie bracket $[-,-]$ here being the 3-bracket
$[-,-,-]$, an alternating trilinear map on a vector space $V$, which
satisfies a natural generalisation of the Jacobi identity (sometimes
referred to as the fundamental identity).  The dynamical fields in the
Bagger--Lambert model take values in $V$ and consist of eight real
bosonic scalars and a fermionic spinor in three dimensions which
transforms as a chiral spinor under the $\fso(8)$ R-symmetry.  There is
also a non-dynamical gauge field which takes values in a Lie subalgebra
of $\fgl(V)$.  The on-shell closure of the supersymmetry transformations
for these fields follows from the fundamental identity.

To obtain the correct equations of motion for the Bagger--Lambert
theory from a lagrangian that is invariant under all the
aforementioned symmetries seems to require the Lie 3-algebra to admit
an invariant inner product. (Following remark 8 in \cite{Lor3Lie} , we
will take this inner product to be non-degenerate.) The signature of
the inner product here determines the relative signs of the kinetic
terms for the scalar and fermion fields in the Bagger--Lambert
lagrangian.  It would therefore be desirable to choose this inner
product to have positive-definite signature, to avoid the occurrence
of states with negative-norm in the quantum theory.  Unfortunately
there are very few such euclidean metric Lie 3-algebras.  Indeed, as
shown in \cite{NagykLie} (see also \cite{GP3Lie, GG3Lie}), they can
always be written as the direct sum of abelian Lie 3-algebras plus
multiple copies of the unique simple euclidean Lie 3-algebra $S_{0,4}$
considered by Bagger and Lambert in their original construction.  The
moduli space for the Bagger--Lambert theory associated with $S_{0,4}$
has been interpreted as describing two M2-branes on a certain M-theory
orbifold in \cite{LambertTong, DistlerBL}.

Thus, in order to find new interacting Bagger--Lambert lagrangians and
despite the possibility of negative-norm states, one is led to
contemplate Lie 3-algebras with an invariant inner product of
indefinite signature.  This idea was pioneered in \cite{GMRBL, BRGTV,
  HIM-M2toD2rev} for a class of Lie 3-algebras (defined by a euclidean
semisimple Lie algebra in two dimensions lower) admitting an inner
product of lorentzian signature. (It was subsequently proved in
\cite{Lor3Lie} that every indecomposable lorentzian Lie 3-algebra
belongs to this class, unless it is the unique simple lorentzian Lie
3-algebra $S_{1,3}$ or one-dimensional.) A feature of these lorentzian
Lie 3-algebras is that their canonical 4-form (built from the
3-algebra structure constants and invariant inner product) has
precisely one leg in only one of the two null directions of the
3-algebra (the remaining three legs are in the directions spanned by
the euclidean Lie algebra).  It is this 4-form which dictates the
structure of the interactions appearing in the Bagger-Lambert
lagrangian.  Consequently the scalar and fermion fields in the other
null direction of the 3-algebra appear only in the free kinetic terms
of the Bagger--Lambert lagrangian here.  This might suggest one could
simply integrate out these components, thus setting the scalar and
fermion components in the complementary null direction to obey their
free field equations.  However, the absence of interaction terms
involving scalars and fermions in one null direction gives rise to an
additional symmetry for their kinetic terms upon shifting them by
constant values.  The gauging of this extra shift symmetry in a
superconformally-invariant manner has been analysed in some detail
recently in \cite{BLSNoGhost,GomisSCFT} where it is found that, after
fixing the gauged symmetry, the resulting lagrangian can be simply
expressed as the sum of the ungauged lagrangian plus a maximally
supersymmetric lagrangian involving the Faddeev--Popov ghosts for the
shift symmetry.  The BRST transformations for this gauge-fixed theory
mix the fields and ghosts in the two lagrangians as expected and the
BRST cohomology is found to be free of negative-norm states.

It would be good to understand whether this remarkable absence of
negative-norm states in the Bagger-Lambert theory for lorentzian Lie
3-algebras persists for inner products of more indefinite signature, or
at least whether one can establish clear 3-algebraic criteria that would
guarantee that the construction noted above could be employed for more
general algebras.  The resulting moduli spaces for such theories and
their possible M-theoretic interpretation might also be of interest.  In
this paper we will begin to address some of these questions.

Given the central rôle played by the Lie 3-algebra in the
Bagger--Lambert theory, we first recast some desirable physical
properties, such as the possibility of decoupling of negative-norm
states and the scale and parity invariances of the model, in 3-algebraic
language.  This then motivates the study of metric Lie 3-algebras in
arbitrary signature subject to some 3-algebraic criteria which can be
explicitly checked for a given class of Lie 3-algebras or else built
\emph{ab initio} into a general construction of such Lie 3-algebras.

Indeed, by analogy with the structure theorem
\cite{MedinaRevoy,FSalgebra} of metric Lie algebras, we will prove that
any metric Lie 3-algebra can be constructed from the one-dimensional and
simple Lie 3-algebras by iterating the operations of orthogonal direct
sum and `double extension' (see Definition~\ref{def:doublext}).
Furthermore, following \cite{Lor3Lie}, we will classify Lie 3-algebras
with inner products of $(2,p)$ signature and find the non-simple
indecomposable ones to fall into one of two distinct classes.  We find
that only one of these two classes describes metric Lie 3-algebras with
a canonical 4-form having no legs in precisely half of the null
directions of the 3-algebra (this is similar to what happened in the
lorentzian case).  For the Bagger--Lambert theory with Lie 3-algebra in
this class, this implies that the scalars and fermions in these two null
directions appear only in the free kinetic terms of the lagrangian,
suggesting that one might be able to remove the negative-norm states
from the physical Hilbert space here also after appropriate fixing of
the gauged shift symmetries as in \cite{BLSNoGhost,GomisSCFT}.  We will
consider the physical properties of such theories and their
corresponding moduli spaces in a forthcoming publication.

We should emphasise that even if the existence of extra shift symmetries
allows one to obtain a positive-definite Hilbert space, in general one
would need to impose extra constraints to make contact with the
effective description of M2-branes.  For instance, one would also like
the effective field theory to have no coupling constant.  This would be
guaranteed provided the Lie 3-algebra admits a suitable outer
automorphism which absorbs the formal coupling dependence by rescaling
the inner product in the Bagger--Lambert lagrangian.  With this in mind,
we compute the Lie algebra of derivations (i.e. the infinitesimal form
of the automorphism group) of each class of metric Lie 3-algebra in
$(2,p)$ signature and highlight, when it exists, the appropriate outer
automorphism that could be used to fix the coupling constant in the
Bagger--Lambert theory.  It would also be desirable for the theory to be
parity-invariant.  This condition is satisfied provided the Lie
3-algebra admits an isometric anti-automorphism (which effectively
reverses the sign of the Lie 3-algebra structure constants in such a way
that it compensates the parity inversion on the M2-branes' worldvolume).
We will find four new classes of $(2,p)$ signature metric Lie 3-algebras
that satisfy all of the above conditions.

We will also make some remarks on how the decoupling of negative-norm
states might work for metric Lie 3-algebras of general indefinite
signature.  The abstract criterion for the existence of additional
shift symmetries in half of the null directions will be that the Lie
3-algebra should admit a maximally isotropic centre.  We show that the
obstruction to having a maximally isotropic centre corresponds roughly
speaking to the existence of a simple ideal among the ingredients of
the metric Lie 3-algebra.  This allows us to give a prescription for
how to construct \emph{in principle} all such metric Lie 3-algebras.

The paper is organised as follows.  In Section~\ref{sec:3-BL} we will
briefly review the main features of the Bagger--Lambert theory from the
perspective of Lie 3-algebras.  In other words, this section will
translate desirable physical properties of the theory into 3-algebraic
criteria.  These will be revisited at the end of the paper in
Section~\ref{sec:no-ghosts} in light of the structural results and
classifications obtained in the paper.  We interpret metricity and
indecomposability in terms of the Bagger--Lambert theory and show why
these properties are desirable.  We then translate physical requirements
such as ghost decoupling, absence of scale and parity invariance into
3-algebraic criteria, which can be easily checked given either explicit
metric Lie 3-algebras as in Section~\ref{sec:2p} or a general
construction scheme as in Section~\ref{sec:metric}.  After this physical
motivation, the paper contains a number of technical algebraic sections.
In Section~\ref{sec:metric} we prove a structure theorem for metric
Lie 3-algebras.  After the observation that every metric Lie 3-algebra
is an orthogonal direct sum of indecomposables, we set out to prove
Theorem~\ref{th:indecmetric3lie}, which says that every indecomposable
metric Lie 3-algebra is either one-dimensional, simple or else it is
obtained by double extending a (not necessarily indecomposable) metric
Lie 3-algebra by either a one-dimensional or simple Lie 3-algebra.  A
simple induction argument then allows us to state
Corollary~\ref{co:metric3lie} describing the class of metric Lie
3-algebras as the class of Lie 3-algebras generated by the
one-dimensional and simple Lie 3-algebras under the operations of double
extension and orthogonal direct sum.  The notion of double extension
appears in Definition~\ref{def:doublext}.  Section~\ref{sec:2p} contains
a classification of indecomposable metric Lie 3-algebras of signature
$(2,p)$, culminating with Theorem~\ref{th:2p}.  They fall into two main
types, which we call \emph{type Ia}, defined in \eqref{eq:type-Ia}, and
\emph{type IIIb}, defined in \eqref{eq:type-IIIb}.  They are
distinguished by the fact that type IIIb possesses a maximally isotropic
centre, whereas type Ia does not.  In fact, type IIIb is a very general
class of metric Lie 3-algebras and we deconstruct it into several
non-isomorphic classes at the end of Section~\ref{sec:type-III}.
Section~\ref{sec:derivations} contains the analysis of the Lie algebra
of derivations of the indecomposable metric Lie 3-algebras found in
Section~\ref{sec:2p}, as well as to the subalgebras of derivations which
preserve (the conformal class of) the inner product.  Finally in
Section~\ref{sec:no-ghosts} we revisit the 3-algebraic criteria obtained
in Section \ref{sec:3-BL} in light of the explicit classification and
structural results obtained in the previous sections.  We focus first on
the absence of ghosts, and using the structure results in
Section~\ref{sec:structure} we indicate how metric Lie 3-algebras
satisfying the ghost-decoupling criterion can be constructed.  For the
$(2,p)$ metric Lie 3-algebras classified in Section~\ref{sec:2p}, we
determine the indecomposable ones which satisfy the ghost decoupling
criterion as well as the absence of a coupling constant and the parity
invariance of the theory.  This results in four classes of
indecomposable $(2,p)$ metric Lie 3-algebras deserving of further
study.

\section*{Acknowledgments}

PdM is supported by a Seggie-Brown Postdoctoral Fellowship of the
School of Mathematics of the University of Edinburgh.

\section{Bagger--Lambert theory in Lie 3-algebraic language}
\label{sec:3-BL}

We will start by summarising how some of the physical properties of the
Bagger--Lambert theory relate to the general structure of metric Lie
3-algebras discussed in \cite{Lor3Lie} and later in this paper.  In
Section~\ref{sec:no-ghosts} we will revisit these properties in light of
our classification of $(2,p)$ signature metric Lie 3-algebras in
Section~\ref{sec:2p} and our structure theorem of
Section~\ref{sec:metric}.

\subsection{Brief review of Bagger--Lambert theory}
\label{sec:brief-BL}

Let us begin by reviewing some details of the Bagger--Lambert theory.
Consider a metric Lie 3-algebra, with 3-bracket $[-,-,-]$ and inner
product $\left<-,-\right>$ of general indefinite signature $(k,k+n)$ and
define a null basis $e_A = (u_i, v_i, e_a)$, with $i=1,...,k$,
$a=1,...,n$, such that $\left<u_i,v_j\right> = \delta_{ij}$,
$\left<u_i,u_j\right> = 0 = \left<v_i,v_j\right>$ and
$\left<e_a,e_b\right> = \delta_{ab}$.  With respect to this basis, the
components of the canonical 4-form for the metric Lie 3-algebra are
$F_{ABCD} := \left< [ e_A , e_B , e_C ] , e_D \right>$ (indices will be
raised and lowered using the invariant 3-algebra metric
$\left<e_A,e_B\right>$).  The fields in the Bagger--Lambert theory have
components $X_I^A$, $\Psi^A$, $(\tilde{A}_\mu)^A{}_B = F^A{}_{BCD}
A_\mu^{CD}$, corresponding respectively to the scalars ($I=1,...,8$ in
the vector of the $\fso(8)$ R-symmetry), fermions and gauge field ($\mu
=0,1,2$ on $\RR^{1,2}$ Minkowski space).  Although the supersymmetry
transformations and equations of motion can be expressed in terms of
$(\tilde{A}_\mu)^A_B$, the lagrangian requires it to be expressed as
above in terms of $A_\mu^{AB}$.  Indeed, recall that the Bagger--Lambert
lagrangian \cite{BL2} can be written
\begin{multline}
  \label{eq:BLLag}
  \eL = -\half \left< D_\mu X_I , D^\mu X_I \right> + \tfrac{i}2 \left< {\bar
      \Psi} , \Gamma^\mu D_\mu \Psi \right> - \tfrac{i}4 \, F_{ABCD}
  ( {\bar \Psi}^A \Gamma^{IJ} \Psi^B ) X^C_I X^D_J\\
  - \tfrac1{12} \left< [X_I,X_J,X_K],[X_I,X_J,X_K]\right> + \eL_{CS}~,
\end{multline}
where
\begin{equation}
  \label{eq:BLCS}
  \eL_{CS} = \half \left( A^{AB} \wedge d \tilde{A}_{AB} +
    \tfrac23 \, A^{AB} \wedge \tilde{A}_{AC} \wedge \tilde{A}^C{}_B
  \right)~,
\end{equation}
and $D_\mu \phi^A = \partial_\mu \phi^A + (\tilde{A}_\mu)^A{}_B
\phi^B$ defines the action on any field $\phi$ valued in $V$ of
the derivative $D$ that is gauge-covariant with respect to
$\tilde{A}^A{}_B$.  The fermion $\Psi^A$ transforms as a Majorana
spinor in $\RR^{1,10}$ subject to the projection $\Gamma_{012} \Psi^A
= -\Psi^A$, where $\Gamma_\mu$ and $\Gamma_I$ denote respectively the
$\RR^{1,2}$ and $\RR^8$ components of the Clifford algebra on
$\RR^{1,10}$ .

The integral of the lagrangian \eqref{eq:BLLag} is invariant under
the supersymmetry transformations
\begin{align*}
  \delta X^A_I &= i\, {\bar \epsilon} \Gamma_I \Psi^A\\
  \delta \Psi^A &= D_\mu X^A_I \Gamma^\mu \Gamma^I \epsilon +
  \tfrac{1}{6} F^A{}_{BCD} X^B_I X^C_J X^D_K \Gamma^{IJK} \epsilon\\
  \delta (\tilde{A}_\mu)^A{}_{B} &= i F^A{}_{BCD} X^C_I {\bar
    \epsilon} \Gamma_\mu \Gamma^I \Psi^D ~,
\end{align*}
where the parameter $\epsilon$ transforms as a Majorana spinor on
$\RR^{1,10}$ subject to the projection $\Gamma_{012} \epsilon =
\epsilon$.  Using the equations of motion obtained from
\eqref{eq:BLLag}, these supersymmetry transformations are found to
close, up to translations in $\RR^{1,2}$ and gauge transformations.

\subsection{Degeneracy implies decoupling}
\label{sec:nondeg}

We have been assuming throughout that the Lie 3-algebra inner product is
nondegenerate.  Recall that the supersymmetry transformations and
equations of motion for the Bagger--Lambert theory do not require the
Lie 3-algebra to admit an inner product at all.  It is the existence of
a lagrangian which realises all the symmetries and gives rise to all the
correct equations of motion appearing in the on-shell closure of
supersymmetry algebra that seems to require that the Lie 3-algebra
should admit an invariant inner product.  As shown in remark 8 of
\cite{Lor3Lie}, if we allowed this invariant inner product to be
degenerate then the corresponding canonical 4-form $F$ can have no legs
along any of the degenerate directions.  Since it is this 4-form which
dictates the structure of interactions in the Bagger--Lambert lagrangian
\eqref{eq:BLLag}, it is clear that only the fields in the nondegenerate
directions can appear in the interactions.  Moreover, even the free
scalar and fermion kinetic terms cannot appear along the degenerate
directions of the inner product.  Hence there is always a consistent
truncation of the theory to the quotient Lie 3-algebra (with
nondegenerate inner product) along the nondegenerate directions.
Therefore, in terms of characterising which kinds of new interactions
can occur in the Bagger--Lambert lagrangian, nothing is lost by assuming
a nondegenerate inner product.

\subsection{Decomposability implies decoupling}
\label{sec:decomp}

By definition, a decomposable metric Lie 3-algebra $V = V_1 \oplus V_2$
can be written as the orthogonal direct sum of two metric Lie 3-algebras
$V_1$ and $V_2$ (with $[V_1,V_2,V] =0$).  This implies that neither the
inner product $\left<-,-\right>$ nor the canonical 4-form $F$ of a
decomposable Lie 3-algebra can ever have `mixed legs' in both $V_1$ and
$V_2$.  Hence the Bagger--Lambert lagrangian and supersymmetry
transformations for $V = V_1 \oplus V_2$ completely decouple in terms of
the individual factors $V_1$ and $V_2$.  Thus, as one would expect, the
indecomposable metric Lie 3-algebras really are the basic building
blocks for the Bagger--Lambert theory.

\subsection{Maximally isotropic centres, shift symmetries and decoupling
  of ghosts}
\label{sec:shifts}

The free kinetic terms for the scalars and fermions
\begin{equation}
  \label{eq:BLfree}
  \sum_{i=1}^k \left( - \partial_\mu X_I^{u_i} \partial^\mu X_I^{v_i} +
    i\, {\bar \Psi}^{u_i} \Gamma^\mu \partial_\mu \Psi^{v_i} \right) -
  \half \partial_\mu X_I^a \partial^\mu X^a_I + \tfrac{i}2\, {\bar \Psi}^a
  \Gamma^\mu \partial_\mu \Psi^a
\end{equation}
are the only terms which do not couple to components $F_{ABCD}$ of
the canonical 4-form in \eqref{eq:BLLag}.  The $(u_i , v_i)$
components correspond to the $2k$ null directions in the Lie
3-algebra and are related to the $k$ kinetic terms with the
`wrong' sign in the lagrangian that need to be exorcised.

From \eqref{eq:BLLag} we see that there are two ways in which the
fields in the Bagger--Lambert theory couple to the components of
$F$ in the interaction terms in the lagrangian; either linearly
via contraction with $F_{ABCD}$ or quadratically via contraction
with $F_{ABCG} F_{DEF}{}^G$.  The first type gives rise
to the $X^2 \Psi^2$, $A X^2$, $A \Psi^2$ interactions and the free
kinetic term for the gauge field.  The second type gives rise to
the $X^6 $, $A^2 X^2$ and $A^3$ interactions.

Clearly a sufficient condition for the existence of extra shift
symmetries for, say, the null components $X^{v_i}$ and $\Psi^{v_i}$ in
\eqref{eq:BLfree} would be that these fields do not appear in any of the
interactions.  If this condition is met for all $k$ of these null
components then a naive counting argument would suggest that
gauge-fixing the theory obtained from gauging these extra shift
symmetries, à la \cite{BLSNoGhost,GomisSCFT}, should have a BRST
cohomology free of negative-norm states.

The question of whether the criterion above is met can be posed at the
level of the Lie 3-algebra as follows.  Of the various terms in the
Bagger--Lambert lagrangian, the most restrictive in terms of satisfying
the condition above comes from the quartic $X^2 \Psi^2$ scalar-fermion
interaction.  For this to not involve any of the null components
$X^{v_i}$ and $\Psi^{v_i}$ requires $F_{v_i ABC} = 0$.  This condition
on the Lie 3-algebra is equivalent to saying that the $v_i$ are central,
whence they span a maximally isotropic subspace of the centre.  Notice
that $F_{v_i ABC} = 0$ further implies $F_{v_i ABG} F_{CDE}{}^G = 0$ and
so, as required, the fields $X^{v_i}$, $\Psi^{v_i}$ in these $k$ null
directions do not appear in any of the interactions (furthermore
$A_\mu^{v_i A}$ does not appear at all in the lagrangian).

It is worth remarking that the criterion above is satisfied for all
indecomposable lorentzian Lie 3-algebras except the unique simple one
$S_{1,3}$.  This follows from Theorem~9 in \cite{Lor3Lie} where, in
the Witt basis $(u,v, e_a )$, the canonical 4-form has only the legs
$F_{uabc} = f_{abc}$ (where $f_{ab}{}^c$ are the structure constants of
a compact semisimple Lie algebra).  On the other hand $S_{1,3}$ has a
non-vanishing $F_{uvab}$ component, thus violating the condition
above.  We will notice a similar structure in Theorem~\ref{th:2p}
below.

\subsection{Conformal automorphisms and the coupling constant}
\label{sec:coupling}

The formal coupling dependence of the interactions in the
Bagger--Lambert theory can be brought to an overall factor
$\frac{1}{\kappa^2}$ multiplying the lagrangian by rescaling the
canonical 4-form $F_{ABCD} \rightarrow \kappa^2 F_{ABCD}$ followed by
the field redefinitions $X^A_I \rightarrow \frac{1}{\kappa} X^A_I$,
$\Psi^A \rightarrow \frac{1}{\kappa} \Psi^A$ and $A_\mu^{AB} \rightarrow
\frac{1}{\kappa^2} A_\mu^{AB}$.  Since the scalars and fermions are
valued in $V$, inspection of all the terms in the lagrangian not
involving the gauge field shows that the overall $\frac{1}{\kappa^2}$
factor can be absorbed for these terms if the Lie 3-algebra admits a
conformal automorphism $\varphi : V \rightarrow V$ with
$\left<\varphi(x) , \varphi(y)\right> = \kappa^2 \left<x,y\right>$ for
all $x,y \in V$.  This would then follow by redefining $X_I \rightarrow
\varphi (X_I)$ and $\Psi \rightarrow \varphi(\Psi)$ in the lagrangian.
This induces the transformation $\tilde{A}_\mu \rightarrow \varphi
\tilde{A}_\mu \varphi^{-1}$ on the gauge field $\tilde{A}_\mu$ that
appears in the covariant derivatives.  The Chern-Simons term then
requires $A_\mu \rightarrow \varphi A_\mu \varphi^t$, which indeed
follows from the definition $(\tilde{A}_\mu)^A{}_B = F^A{}_{BCD}
A_\mu^{CD}$, in order to get the correct scaling for all the terms
involving the gauge field in the Bagger--Lambert lagrangian.

Let us recall how this works for the case of indecomposable
lorentzian Lie 3-algebras, as noted in \cite{BRGTV,
HIM-M2toD2rev}, using the language of Proposition~11 in
\cite{Lor3Lie}.  Relative to the Witt basis $(u,v, e_a )$ noted
above, the appropriate conformal automorphism maps $\varphi (u) =
\beta u$, $\varphi (v) = \beta^{-3} v$ and $\varphi (e_a) =
\beta^{-1} e_a$, which rescales the inner product by a factor of
$\beta^{-2}$ and so absorbs the coupling if we take $\beta =
\kappa^{-1}$.  To be completely explicit, $X_I \rightarrow
\varphi( X_I )$ transforms the components of the scalars as $X^u_I
\rightarrow \kappa^{-1} X^u_I$, $X^v_I \rightarrow \kappa^{3}
X^v_I$, $X^a_I \rightarrow \kappa X^a_I$ (and likewise for the
fermions), whereas $A_\mu \rightarrow \varphi A_\mu \varphi^t$
transforms the components of the gauge field as $A_\mu^{uv}
\rightarrow \kappa^2 A_\mu^{uv}$, $A_\mu^{ua} \rightarrow
A_\mu^{ua}$, $A_\mu^{va} \rightarrow \kappa^4 A_\mu^{va}$,
$A_\mu^{ab} \rightarrow \kappa^2 A_\mu^{ab}$. (Recall though that
the central components $A_\mu^{vA}$ do not appear in the
lagrangian.) The components $B_a := \half f_{abc} A^{bc}$ and $A^a
:= A^{ua}$ of the gauge field appear in the Chern-Simons term of
\eqref{eq:BLLag} as a so-called BF term (with $F$ being the field
strength of gauge field $A$ here).  This automorphism then matches
equation (2.30) in \cite{BRGTV} if we identify their $+,-$ and $g$
with our $u,v$ and $\kappa$.

In Section~\ref{sec:no-ghosts} we will present a similar story for
several of the classes of metric Lie 3-algebras in $(2,p)$ signature in
Section~\ref{sec:2p}.

\subsection{Isometric anti-automorphisms and parity invariance}
\label{sec:parity}

Recall \cite{SchwarzChernSimons} that the effective field theory on the
worldvolume of M2-branes is expected to be invariant under the $\ZZ_2$
symmetry generated by a parity inversion.  Despite the existence of a
Chern-Simons term, this condition is satisfied for the Bagger--Lambert
theory based on the euclidean simple Lie 3-algebra $S_{0,4}$ (see
\cite{BandresCS, VanRaamsdonkBL}) and for the class of indecomposable
lorentzian Lie 3-algebras in \cite{BRGTV, HIM-M2toD2rev}.

Following \cite{BL2,BandresCS}, let us define a parity inversion
to map the $\RR^{1,2}$ coordinates $(x^0,x^1,x^2) \rightarrow
(x^0,x^1,-x^2)$ (thus reversing the orientation on $\RR^{1,2}$)
and mapping spinors $\Psi \rightarrow \Gamma_2 \Psi$.  This implies
that the spinor bilinear terms ${\bar \Psi}^A \Gamma^\mu
\partial_\mu \Psi^B$ and ${\bar \Psi}^A \Gamma^{IJ} \Psi^B$ in the Bagger--Lambert Lagrangian
\eqref{eq:BLLag} are respectively even and odd under this parity
transformation.  The structure of covariant derivatives in
\eqref{eq:BLLag} further implies that $\tilde{A}_0$, $\tilde{A}_1$
should be parity-even whilst $\tilde{A}_2$ is parity-odd.  Given
these parity assignments, inspection of the terms in the
Bagger--Lambert Lagrangian \eqref{eq:BLLag} leads one to deduce
that a sufficient condition for invariance is that the Lie
3-algebra admits an isometric anti-automorphism, i.e. a linear map
$\gamma : V \rightarrow V$ obeying $\left< \gamma x , \gamma y
\right> = \left<x,y\right>$ and $[\gamma x , \gamma y , \gamma z ]
= - \gamma [x,y,z]$ for all $x,y,z$ in $V$. (This results in
effectively reversing the sign of all the structure constants in
the Lie 3-algebra, a condition already noted in \cite{BL2} for
parity-invariance.) The transformation $\gamma$ maps $X_I
\rightarrow \gamma X_I$ and $\tilde{A}_\mu \rightarrow \gamma
\tilde{A}_\mu \gamma^{-1}$, which implies $A_\mu \rightarrow -
\gamma A_\mu \gamma^t$.  Combining this transformation with the
action of parity on the fields thus leaves \eqref{eq:BLLag}
invariant.  Hence the criterion of parity-invariance can be reduced
to the existence of an isometric anti-automorphism of the Lie
3-algebra.  Notice that the composition of any two isometric
anti-automorphisms is an isometric automorphism, i.e. a symmetry
of \eqref{eq:BLLag} on its own.  Thus the generator of the
isometric anti-automorphism needed for parity-invariance is
essentially unique modulo a symmetry of the Lagrangian.

For the euclidean case $S_{0,4}$, defined with respect to an
orthonormal basis $(e_1,e_2,e_3,e_4)$, the appropriate isometric
anti-automorphism can be taken to be the map $\gamma
(e_1,e_2,e_3,e_4) = (-e_1,-e_2,-e_3,e_4)$, as was clarified in
\cite{VanRaamsdonkBL}.  For the lorentzian case discussed in
\cite{BRGTV, HIM-M2toD2rev}, relative to the Witt basis $(u,v, e_a
)$ we have defined above, the appropriate anti-automorphism maps
$\gamma (u,v, e_a ) = (u,v, -e_a )$.  Reading off the corresponding
transformations of the components of the fields above, one finds
that $X^a_I$, $\Psi^a$, $A_\mu^{uv}$ and $A_\mu^{ab}$ are odd
under the action of $\gamma$ whilst all the other components are
even (in agreement with \cite{BRGTV}).

\subsection{Summary}
\label{sec:summary}

In summary, we are interested in classifying or understanding how to
construct \emph{indecomposable} \emph{metric} Lie 3-algebras admitting a
\emph{maximally isotropic centre}, \emph{conformal derivations} and
\emph{isometric anti-auto\-morphisms}.  In the next sections we will
prove a structure theorem for metric Lie 3-algebras, classify those with
signature $(2,p)$ and study their Lie algebras of (conformal)
derivations.  Finally in Section~\ref{sec:no-ghosts}, we will revisit
these conditions in light of the above results.

\section{Metric Lie $3$-algebras and double extensions}
\label{sec:metric}

Recall that a (finite-dimensional, real) Lie 3-algebra consists of a
finite-dimensional real vector space $V$ together with a linear map
$\Phi: \Lambda^3 V \to V$, denoted simply as a 3-bracket, obeying a
generalisation of the Jacobi identity.  To define it, let us recall
that an endomorphism $D\in \End V$ is said to be a \textbf{derivation}
if
\begin{equation*}
  D[x_1,x_2,x_3] = [Dx_1,x_2,x_3] + [x_1,Dx_2,x_3] + [x_1,x_2,Dx_3]~,
\end{equation*}
for all $x_i \in V$.  Then $(V,\Phi)$ defines a \textbf{Lie
  3-algebra} if the endomorphisms $\ad_{x_1,x_2} \in \End
V$, defined by $\ad_{x_1,x_2} y = [x_1,x_2,y]$, are derivations:
\begin{equation}
  \label{eq:3-Jacobi}
  [x_1,x_2,[y_1,y_2,y_3]] =
  [[x_1,x_2,y_1],y_2,y_3] +   [y_1,[x_1,x_2,y_2],y_3] +
  [y_1,y_2,[x_1,x_2,y_3]]~,
\end{equation}
for all $y_i\in V$.  We call it the 3-Jacobi identity or, in the present
context, the fundamental identity.  The vector space of derivations is a
Lie subalgebra of $\fgl(V)$ denoted $\Der V$.  The derivations
$\ad_{x_1,x_2} \in \Der V$ span the ideal $\ad V \lhd \Der V$ consisting
of \textbf{inner derivations}.

We recall that a metric Lie 3-algebra is a triple $(V,\Phi,b)$
consisting of a finite-dimensional real Lie 3-algebra $(V,\Phi)$
together with a nondegenerate symmetric bilinear form $b:S^2 V \to \RR$,
denoted simply by $\left<-,-\right>$, subject to the invariance
condition of the inner product
\begin{equation}
  \label{eq:adinvariance}
  \left<[x_1,x_2,y_1],y_2\right> = -
  \left<[x_1,x_2,y_2],y_1\right>~,
\end{equation}
for all $x_i, y_i \in V$.

Given two metric Lie 3-algebras $(V_1,\Phi_1,b_1)$ and
$(V_2,\Phi_2,b_2)$, we may form their \textbf{orthogonal direct sum}
$(V_1\oplus V_2,\Phi_1\oplus \Phi_2, b_1 \oplus b_2)$, by declaring
that
\begin{align*}
  [x_1,x_2,y] = 0 \qquad\text{and}\qquad
  \left<x_1,x_2\right> = 0~,
\end{align*}
for all $x_i\in V_i$ and all $y\in V_1 \oplus V_2$.  The resulting
object is again a metric Lie 3-algebra.  A metric Lie 3-algebra is
said to be \textbf{indecomposable} if it is not isomorphic to an
orthogonal direct sum of metric Lie 3-algebras $(V_1\oplus 
V_2, \Phi_1\oplus \Phi_2, b_1\oplus b_2)$ with $\dim V_i > 0$.
In order to classify the metric Lie 3-algebras, it is clearly enough
to classify the indecomposable ones.  In this section we will prove a
structure theorem for indecomposable Lie 3-algebras.  We will prove that
they are constructed from the simple and the one-dimensional Lie
3-algebras by iterating two constructions: the orthogonal direct sum
just defined and the ``double extension'' to be defined below.

\subsection{Basic notions and notation}
\label{sec:basic}

From now on let $(V,\Phi)$ be a Lie 3-algebra.  Given subspaces $W_i
\subset V$, we will let $[W_1 W_2 W_3]$ denote the subspace of $V$
consisting of elements $[w_1,w_2,w_3] \in V$, where $w_i \in W_i$.

We will use freely the notions of subalgebra, ideal and homomorphisms
as reviewed in \cite{Lor3Lie}.  In particular a \textbf{subalgebra}
$W < V$ is a subspace $W \subset V$ such that $[W W W] \subset W$,
whereas an \textbf{ideal} $I \lhd V$ is a subspace $I \subset V$ such
that $[I V V] \subset I$.  A linear map $\phi: V_1 \to V_2$
between Lie 3-algebras is a \textbf{homomorphism} if
$\phi[x_1,x_2,x_3] = [\phi(x_1),\phi(x_2),\phi(x_3)]$, for all $x_i\in
V_1$.  An \textbf{isomorphism} is a bijective homomorphism.  There is
a one-to-one correspondence between ideals and homomorphisms and all
the standard theorems hold.  In particular, intersection and sums of
ideals are ideals.  An ideal $I\lhd V$ is said to be \textbf{minimal}
if any other ideal $J\lhd V$ contained in $I$ is either $0$ or $I$.
Dually, an ideal $I\lhd V$ is said to be \textbf{maximal} if any other
ideal $J\lhd V$ containing $I$ is either $V$ and $I$.  A Lie
3-algebra is said to be \textbf{simple} if it has no proper ideals
and $\dim V > 1$.

\begin{lemma}\label{le:simplequot}
  If $I\lhd V$ is a maximal ideal, then $V/I$ is simple or
  one-dimensional.
\end{lemma}

Simple Lie 3-algebras have been classified.

\begin{theorem}[{\cite[§3]{LingSimple}}]\label{th:simple}
  A simple real Lie 3-algebra is isomorphic to one of the
  Lie 3-algebras defined, relative to a basis
  $(\be_i)_{i=1,2,3,4}$, by
  \begin{equation}
    \label{eq:simple-3-Lie}
    [\be_1,\dots, \widehat{\be_i},\dots, \be_4] = (-1)^i \varepsilon_i
    \be_i~,
  \end{equation}
  where a hat denotes omission and where the $\varepsilon_i$ are
  signs.
\end{theorem}

It is plain to see that simple real Lie 3-algebras admit invariant inner
products of any signature.  Indeed, the Lie 3-algebra in
\eqref{eq:simple-3-Lie} leaves invariant the diagonal inner product with
entries $(\varepsilon_1, \dots, \varepsilon_4)$.  We will let $S_{p,q}$
denote the simple metric Lie 3-algebra with signature $(p,q)$.  There
are, up to homothety, precisely three: $S_{0,4}$, $S_{1,3}$ and
$S_{2,2}$, corresponding to euclidean, lorentzian and split signatures,
respectively.  The Lie 3-bracket of $S_{p,q}$ is given relative to a
basis $(\be_1,\be_2,\be_3,\be_4)$ by \eqref{eq:simple-3-Lie} where the
signs $\varepsilon_i$ are given by $(++++)$ for $S_{0,4}$, $(-+++)$ for
$S_{1,3}$ and $(--++)$ for $S_{2,2}$.  Implicit in the notation is a
choice of invariant inner product, which is given by
$\left<\be_i,\be_j\right> = \lambda \varepsilon_i \delta_{ij}$, for some
$\lambda>0$.

Complementary to the notion of (semi)simplicity is that of solvability.
As shown by Kasymov \cite{Kasymov}, there are more than one notion
of solvability for Lie 3-algebras.  However we will use here 
the original notion introduced by Filippov \cite{Filippov}.  Let $I
\lhd V$ be an ideal.  We define inductively a sequence of ideals
\begin{equation}
  \label{eq:sequence}
  I^{(0)} = I \qquad\text{and}\qquad
  I^{(k+1)} = [I^{(k)} I^{(k)} I^{(k)}] \subset I^{(k)}~.
\end{equation}
We say that $I$ is \textbf{solvable} if $I^{(s)}=0$ for some $s$, and
we say that $V$ is \textbf{solvable} if it is solvable as an ideal of
itself.  If $I,J\lhd V$ are solvable ideals, so is their sum $I + J$,
leading to the notion of a maximal solvable ideal $\Rad V$, known as
the \textbf{radical} of $V$.  A Lie 3-algebra $V$ is said to be
\textbf{semisimple} if $\Rad V=0$.  Ling \cite{LingSimple} showed that
a semisimple Lie 3-algebra is isomorphic to the direct sum of its
simple ideals.  The following result is due to Filippov \cite{Filippov}
and can be paraphrased as saying that the radical is a
\emph{characteristic} ideal.

\begin{theorem}[{\cite[Theorem~1]{Filippov}}]\label{th:RadChar}
  Let $V$ be a Lie 3-algebra.  Then $D\Rad V \subset \Rad V$ for
  every derivation $D \in \Der V$.
\end{theorem}

We say that a subalgebra $L < V$ is a \textbf{Levi subalgebra} if $V =
L \oplus \Rad V$ as vector spaces.  Ling showed that, as in the theory
of Lie algebras, Lie 3-algebras admit a Levi decomposition.

\begin{theorem}[{\cite[Theorem~4.1]{LingSimple}}]\label{th:Levi}
  Let $V$ be a Lie 3-algebra.  Then $V$ admits a Levi subalgebra.
\end{theorem}

A further result of Ling's which we shall need is the following.

\begin{theorem}[{\cite[§2]{LingSimple}}]\label{th:emissingling}
  Let $V$ be a Lie 3-algebra.  Then $V$ is semisimple if and only if
  the Lie algebra $\ad V$ of inner derivations is semisimple and all
  derivations are inner, so that $\Der V = \ad V$.
\end{theorem}




In turn this allows us to prove the following useful result.

\begin{proposition}\label{pr:exactness}
  Let $0 \to A \to B \to \Cbar \to 0$ be an exact sequence of Lie
  3-algebras.  If $A$ and $\Cbar$ are semisimple, then so is $B$.
\end{proposition}

\begin{proof}
  Since $A$ is semisimple, Theorem~\ref{th:emissingling} says that
  $\ad A$ is semisimple.  $B$ is a representation of $\ad A$, hence
  fully reducible.  Since $A$ is an $\ad A$-subrepresentation of of
  $B$, we have $B = A \oplus C$, where $C$ is a complementary $\ad
  A$-subrepresentation.  Since $A\lhd B$ is an ideal (being the kernel
  of a homomorphism), $\ad A (C) = 0$, whence $[A A C]=0$.

  The subspace $C$ is actually a subalgebra, since the component
  $[C C C]_A$ of $[C C C]$ along $A$ is $\ad A$-invariant by
  the 3-Jacobi identity and the fact that $C$ is $\ad A$-invariant.
  This means that $[C C C]_A$ is central in $A$, but $A$ is
  semisimple, whence it must vanish.  Hence, $[C C C]\subset C$.
  Since the projection $B \to \Cbar$ maps $C$ isomorphically to
  $\Cbar$, we see that this isomorphism is one of Lie 3-algebras,
  hence $C<B$ is semisimple and indeed $[C C C]=C$.

  It remains to show that $[A C C]=0$.  For $c_i\in C$, the restriction
  of $\ad_{c_1,c_2}$ to $A$ is a derivation of $A$, which belongs to
  $\ad A$ since for $A$ semisimple, all derivations are inner.  Since
  $C$ is $\ad A$-invariant, the 3-Jacobi identity says that $\ad_{c_1,c_2}$ is
  $\ad A$-invariant, whence it belongs to the centre of $\ad A$.
  However $\ad A$ is semisimple  and hence its centre is $0$.

  In summary, $B = A \oplus C$ as a Lie 3-algebra.    Since $A$ and $C$
  are semisimple and hence a sum of simple ideals, so is $B$.
\end{proof}

A useful notion that we will need is that of a representation of a Lie
3-algebra.  A \textbf{representation} of Lie
3-algebra $V$ on a vector space $W$ is a Lie 3-algebra structure
on the direct sum $V \oplus W$ satisfying the following three
properties:
\begin{enumerate}
\item the natural embedding $V \to V \oplus W$ sending $v \mapsto
  (v,0)$ is a Lie 3-algebra homomorphism, so that $[V V V]
  \subset V$ is the original 3-bracket on $V$;
\item $[V V W] \subset W$; and
\item $[V W W]=0$.
\end{enumerate}
The second of the above conditions says that we have a map $\ad V \to
\End W$ from inner derivations of $V$ to linear transformations on
$W$.  The 3-Jacobi identity for $V\oplus W$ says that this map is a
representation of the Lie algebra $\ad V$.  Viceversa, any
representation $\ad V \to \End W$ defines a Lie 3-algebra structure on
$V \oplus W$ extending the Lie 3-algebra structure of $V$ and
demanding that $[V W W]=0$.  Taking $W=V$ gives rise to the
\textbf{adjoint representation}, whereas taking $W = V^*$ gives rise
to the \textbf{coadjoint representation}, where if $\alpha \in V^*$
then
\begin{equation}
  \label{eq:coadjoint}
  [v_1,v_2,\alpha] = \beta \in V^* \qquad\text{where}\qquad
  \beta(v) = - \alpha\left([v_1,v_2,v]\right)~,
\end{equation}
for all $v,v_i\in V$.

Let us now introduce an inner product, so that $(V,\Phi,b)$ is a
metric Lie 3-algebra.

If $W \subset V$ is any subspace, we define
\begin{equation*}
  W^\perp = \left\{v \in V\middle | \left<v,w\right>=0~,\forall w\in
      W\right\}~.
\end{equation*}
Notice that $(W^\perp)^\perp = W$.  We say that $W$ is
\textbf{nondegenerate}, if $W \cap W^\perp = 0$, whence $V = W \oplus
W^\perp$; \textbf{isotropic}, if $W \subset W^\perp$; and
\textbf{coisotropic}, if $W \supset W^\perp$.  Of course, in
positive-definite signature, all subspaces are nondegenerate.

An equivalent criterion for decomposability is the existence of a proper
nondegenerate ideal: for if $I\lhd V$ is nondegenerate, $V = I \oplus
I^\perp$ is an orthogonal direct sum of ideals.  For the proofs of the
following results, the reader is asked to consult \cite[§2.2]{Lor3Lie}.

\begin{lemma}\label{le:coisoquot}
  Let $I\lhd V$ be a coisotropic ideal of a metric Lie 3-algebra.
  Then $I/I^\perp$ is a metric Lie 3-algebra.
\end{lemma}

If $I\lhd V$ is an ideal, the \textbf{centraliser} $Z(I)$ is defined by
the condition $[Z(I)I\,V]=0$.  Taking $V$ as an ideal of itself, we
arrive at the \textbf{centre} $Z(V)$ of $V$.

\begin{lemma}\label{le:centreperp}
  Let $V$ be a metric Lie 3-algebra.  Then the centre is the
  orthogonal subspace to the derived ideal; that is,
  $[V V V]=Z(V)^\perp$.
\end{lemma}

\begin{proposition}\label{pr:ideals}
  Let $V$ be a metric Lie 3-algebra and $I \lhd V$ be an ideal.
  Then
  \begin{enumerate}
  \item $I^\perp \lhd V$ is also an ideal;
  \item $I^\perp\lhd Z(I)$; and
  \item if $I$ is minimal then $I^\perp$ is maximal.
  \end{enumerate}
\end{proposition}

\subsection{Structure of metric Lie 3-algebras}
\label{sec:structure}

We now investigate the structure of metric Lie 3-algebras.  If a Lie
3-algebra $V$ is not simple or one-dimensional, then it has a proper
ideal and hence a minimal ideal.  Let $I\lhd V$ be a minimal ideal of
a metric Lie 3-algebra.  Then $I \cap I^\perp$, being an ideal
contained in $I$, is either $0$ or $I$.  In other words, minimal
ideals are either nondegenerate or isotropic.  If nondegenerate, $V =
I \oplus I^\perp$ is decomposable.  Therefore if $V$ is
indecomposable, $I$ is isotropic.  Moreover, by
Proposition~\ref{pr:ideals} (2), $I$ is abelian and furthermore,
because $I$ is isotropic, $[I\, I\, V]=0$.

It follows that if $V$ is euclidean and indecomposable, it is either
one-dimensional or simple, whence of the form \eqref{eq:simple-3-Lie}
with all $\varepsilon_i=1$.  This result, originally due to Nagy
\cite{NagykLie} (see also \cite{GP3Lie,GG3Lie}), was conjectured in
\cite{FOPPluecker}.

Let $V$ be an indecomposable metric Lie 3-algebra.  Then $V$ is either
simple, one-dimensional (provided the index of the inner product is
$<2$) or possesses an isotropic proper minimal ideal $I$ which obeys
$[I\, I\, V]=0$.  The perpendicular ideal $I^\perp$ is maximal and hence
by Lemma~\ref{le:simplequot}, $\Ubar := V/I^\perp$ is simple or
one-dimensional, whereas by Lemma~\ref{le:coisoquot}, $\Wbar
:=I^\perp/I$ is a metric Lie 3-algebra.  The inner product on $V$
induces a nondegenerate pairing $g: \Ubar \otimes I \to \RR$.  Indeed,
let $[u] = u + I^\perp \in \Ubar$ and $v\in I$.  Then we define
$g([u],v) = \left<u,v\right>$, which is clearly independent of the
coset representative for $[u]$.  In particular, $I \cong \Ubar^*$ is
either one- or 4-dimensional.  If the signature of the inner product
of $\Wbar$ is $(p,q)$, that of $V$ is $(p+r,q+r)$ where $r = \dim I =
\dim \Ubar$.

There are two possibilities for $\Ubar$: either it is one-dimensional or
else it is simple.  We will treat both cases separately.

\subsubsection{$\Ubar$ is one-dimensional}
\label{sec:U1diml}

If the quotient Lie 3-algebra $\Ubar=V/I^\perp$ is one-dimensional,
so is the minimal ideal $I$.  Let $u \in V$ be such that $u \not\in
I^\perp$, whence its image in $\Ubar$ generates it.  Because $I \cong
\Ubar^*$ is induced by the inner product, there is $v \in I$ such that
$\left<u,v\right> = 1$.  The subspace spanned by $u$ and $v$ is
therefore nondegenerate, and hence as a vector space we have an
orthogonal decomposition $V = \RR(u,v) \oplus W$, where $W$ is the
perpendicular complement of $\RR(u,v)$.  It is clear that $W \subset
I^\perp$, and that $I^\perp = I \oplus W$ as a vector space.  Indeed,
the projection $I^\perp \to \Wbar$ maps $W$ isomorphically onto
$\Wbar$.

From Proposition~\ref{pr:ideals} (2), it is immediate that
$[u,v,x]=0=[v,w_1,w_2]$, for all $w_i \in W$, whence $v$ is central.
Metricity then implies that the only nonzero 3-brackets take the form
\begin{equation}
  \label{eq:1dim-3-brackets}
  \begin{aligned}[m]
    [u,w_1,w_2] &= [w_1,w_2]\\
    [w_1,w_2,w_3] &= - \left<[w_1,w_2],w_3\right> v +
    [w_1,w_2,w_3]_W~,
  \end{aligned}
\end{equation}
which defines $[w_1,w_2]$ and $[w_1,w_2,w_3]_W$ and where $w_i\in W$.
The 3-Jacobi identity is equivalent to the following two conditions:
\begin{enumerate}
\item $[w_1,w_2]$ defines a Lie algebra structure on $W$, which
  leaves the inner product invariant due to the skewsymmetry of
  $\left<[w_1,w_2],w_3\right>$; and
\item $[w_1,w_2,w_3]_W$ defines a metric Lie 3-algebra
  structure on $W$ which is invariant under the Lie algebra structure.
\end{enumerate}

We will see below that this says that $V$ is the \emph{double
  extension} of the metric Lie 3-algebra $W$ by the one-dimensional
Lie 3-algebra $\bar U$.

\subsubsection{$\Ubar$ is simple}
\label{sec:u-simple}

Consider $I^\perp$ as a Lie 3-algebra in its own right and let $R=\Rad
I^\perp$ denote its radical.  By Theorem~\ref{th:Levi}, $I^\perp$ admits
a Levi subalgebra $L<I^\perp$.  Since $I^\perp\lhd V$ and $R\lhd
I^\perp$ is a characteristic ideal, $R \lhd V$.  Indeed, for all $x_i\in
V$, $\ad_{x_1,x_2}$ is a derivation of $I^\perp$ (since $I^\perp \lhd
V$) and by Theorem~\ref{th:RadChar}, it preserves $R$.  Let $M = V/R$.
Notice that
\begin{equation*}
  \Ubar = V/I^\perp \cong (V/R)/(I^\perp/R) = M/L~.
\end{equation*}
Since $L$ and $\Ubar$ are semisimple, Proposition~\ref{pr:exactness}
says that so is $M$ and moreover that $M \cong L \oplus \Ubar$.  This
means that $R$ is also the radical of $V$, whence $M$ is a Levi factor
of $V$.  This discussion is summarised by the following commutative
diagram with exact rows and columns:
\begin{equation*}
  \begin{CD}
    @.    0     @.    0     @.    @.  \\
    @.  @VVV         @VVV    @.    @.  \\
    @.    R    @=     R      @.    @.  \\
    @.  @VVV         @VVV    @.    @.  \\
 0 @>>>  I^\perp @>>> V  @>>> \Ubar @>>> 0\\
 @.      @VVV        @VVV      @|   @.\\
   0 @>>>  L    @>>> M  @>>> \Ubar @>>> 0\\
    @.  @VVV         @VVV    @.    @.  \\
    @.    0     @.    0     @.    @.
  \end{CD}
\end{equation*}

The map $M \to \Ubar$ admits a section, so that $M$ has a subalgebra
$\widetilde U$ isomorphic to $\Ubar$ and such that $M = \widetilde U
\oplus L$.  Then the vertical map $V \to M$ also admits a section,
whence there is a subalgebra $U < V$ isomorphic to $\Ubar$ such that
$V = I^\perp \oplus U$ (as vector space).  Furthermore, the inner
product on $V$ pairs $I$ and $U$ nondegenerately, whence $I \oplus U$
is a nondegenerate subspace.  Let $W$ denote its perpendicular
complement, whence $V = W \oplus I \oplus U$.  Clearly $I^\perp = W
\oplus I$, whence the canonical projection $I^\perp \to \Wbar$ maps
$W$ isomorphically onto $\Wbar$.

Let us now write the possible 3-brackets for $V = W \oplus I \oplus U$.
First of all, by Proposition~\ref{pr:ideals} (2), $[V,I^\perp,I]=0$.
Since $U<V$, $[U U U]\subset U$ and since $I$ is an ideal,
$[U U I]\subset I$.  Similarly, since $W \subset I^\perp$ and $I^\perp
\lhd V$ is an ideal, $[W W W]\subset W \oplus I$.  We write this as
\begin{equation*}
  [w_1,w_2,w_3] := [w_1,w_2,w_3]_W + \varphi(w_1,w_2,w_3)~,
\end{equation*}
where $[w_1,w_2,w_3]_W$ defines an 3-bracket on $W$, which is
isomorphic to the Lie 3-bracket of $\Wbar = I^\perp/I$, and
$\varphi:\Lambda^3W \to I$ is to be understood as an abelian
extension.  It remains to understand $[U W W]$ and $[U U W]$.
First of all, we notice that because $W \subset I^\perp$ which is an
ideal, \emph{a priori} $[U W W] \subset W \oplus I$ and $[U U W] \subset
W \oplus I$.   However,
\begin{equation*}
  \left<[U U W],U\right> = - \left<[U U U], W\right> = 0
\end{equation*}
whence the component of $[U U W]$ along $I$ vanishes, so that
$[U U W]\subset W$.  Furthermore, the 3-Jacobi identity makes $W$ into
an $\ad U$-representation and $\varphi$ into an $\ad U$-equivariant
map.

Similarly,
\begin{equation*}
  \left<[U W W],W\right> = - \left<[W W W], U\right>~,
\end{equation*}
whence the $W$ component of $[U W W]$ is determined by the map
$\varphi$ defined above; whereas the $I$ component
\begin{equation*}
  \left<[U W W],U\right> =  \left<[U U W], W\right>
\end{equation*}
is thus determined by the action of $\ad U$ on $W$.

In summary, we have the following nonzero 3-brackets
\begin{equation*}
  [U U U] \subset U \qquad
  [U U I] \subset I \qquad 
  [U U W] \subset W \qquad
  [U W W] \subset W \oplus I \qquad
  [W W W] \subset W \oplus I~,
\end{equation*}
which we will proceed to explain.  The first bracket is simply the
fact that $U<V$ is a subalgebra, whereas the second makes $I$ into a
representation of $U$.  In fact, $I\cong U^*$ is the coadjoint
representation \eqref{eq:coadjoint}.  The third bracket defines an
action of $\ad U$ on $W$ and this also determines the $I$-component of
the fourth bracket.  The $W$-component of the fourth bracket is
determined by the $I$-component of the last bracket.  The last bracket
defines a Lie 3-algebra structure on $W \oplus I$, which is an abelian
extension of the Lie 3-algebra structure on $\Wbar$ by a ``cocycle''
$\varphi: \Lambda^3 W \to I$.  The inner product is such that
$\left<W,W\right>$ and $\left<U,I\right>$ are nondegenerate and the
only other nonzero inner product is $\left<U,U\right>$ which can be
\emph{any} $\ad U$-invariant symmetric bilinear form on $U$, not
necessarily nondegenerate.

Similarly to the case when $\Ubar$ is one-dimensional, we will
interpret $V$ as the \emph{double extension} of the metric Lie
3-algebra $\Wbar$ by the simple Lie 3-algebra $U$.

More generally we have the following definition.

\begin{definition}\label{def:doublext}
  Let $W$ be a metric Lie 3-algebra and let $U$ be a Lie
  3-algebra.  Then by the \textbf{double extension of $W$ by $U$} we
  mean the metric Lie 3-algebra on the vector space $W \oplus U
  \oplus U^*$ with the following nonzero 3-brackets:
  \begin{itemize}
  \item $[UUU] \subset U$ being the bracket of the Lie 3-algebra $U$;
  \item $[UUU^*]\subset U^*$ being the coadjoint action of $\ad U$ on $U^*$;
  \item $[UUW] \subset W$ being the action of $\ad U$ on $W$;
  \item $[UWW] \subset W \oplus U^*$, where the $U^*$ component is
    related to the previous bracket by
    \begin{equation*}
      \left<[u_1,w_1,w_2],u_2\right>  =
      \left<[u_1,u_2,w_1],w_2\right>~.
    \end{equation*}
  \item $[WWW] \subset W \oplus U^*$, where the $W$ component is the
    bracket of the Lie 3-algebra $W$ and the $U^*$ component is
    related to the $W$ component of the previous bracket by
    \begin{equation*}
      \left<[w_1,w_2,w_3],u_1\right>  =
      \left<u_1,w_1,w_2],w_3\right>~.
    \end{equation*}
  \end{itemize}
  These brackets are subject to the 3-Jacobi identity.  Two of these
  identities can be interpreted as saying that the bracket $[UUW]
  \subset W$ defines a Lie algebra homomorphism $\ad U \to \Der^0W$,
  where $\Der^0W$ is the Lie algebra of skewsymmetric derivations of
  the Lie 3-algebra $W$, whereas the map $\Lambda^3W \to U^*$
  defining the $U^*$ component of the $[WWW]$ bracket is $\ad
  U$-equivariant.  We have not found similarly transparent
  interpretations for the other Jacobi identities.  The above
  3-brackets leave invariant the inner product on $V$ with components
  \begin{itemize}
  \item $\left<W,W\right>$, being the inner product on the metric Lie
    3-algebra $W$;
  \item $\left<U,U^*\right>$, being the natural dual pairing; and
  \item $\left<U,U\right>$, being any $\ad U$-invariant symmetric
    bilinear form.
  \end{itemize}
\end{definition}

\begin{remark}
  It can be shown that if $U$ is simple and $[U U W]=0$ then the
  resulting double extension is decomposable.  Indeed, if $[U U W]=0$,
  then by Jacobis the $U^*$ component in $[W W W]$ would have to be
  invariant under $\ad U$.  If $U$ is simple, then this means that
  this component is absent, whence $W$ would be a subalgebra and
  indeed an ideal since the $U^*$ component in $[U W W]$ is also
  absent.  But $W$ is nondegenerate, whence it decomposes $V$.
\end{remark}

In summary, we have proved the following result.

\begin{theorem}\label{th:indecmetric3lie}
  Every indecomposable metric Lie 3-algebra $V$ is either
  one-dimensional, simple or else it is the double extension of a
  metric Lie 3-algebra $W$ by a one-dimensional or simple Lie
  3-algebra $U$.
\end{theorem}

Any metric Lie 3-algebra will be an orthogonal direct sum of
indecomposables, each one being either one-dimensional, simple or
a double extension of a metric Lie 3-algebra, which itself is an
orthogonal direct sum of indecomposables of strictly lower dimension.
Continuing in this way, we arrive at the following characterisation.

\begin{corollary}\label{co:metric3lie}
  The class of metric Lie 3-algebras is generated by the simple and
  one-dimensional Lie 3-algebras under the operations of orthogonal
  direct sum and double extension.
\end{corollary}

It is clear that the subclass of euclidean metric Lie 3-algebras is
generated by the simple and one-dimensional euclidean Lie 3-algebras
under orthogonal direct sum, since double extension always incurs in
indefinite signature.  The lorentzian indecomposables admit at most
one double extension by a one-dimensional Lie 3-algebra and are easy
to classify \cite{Lor3Lie}.  The indecomposables of signature $(2,p)$
will admit at most two double extensions by one-dimensional Lie
3-algebras.  We will find that there are three kinds of such metric
Lie algebras: a simple Lie 3-algebra, one which can be written as a
double extension and one which is the result of iterating two double
extensions.

\section{Metric Lie 3-algebras with signature $(2,p)$}
\label{sec:2p}

In \cite{Lor3Lie} we classified the lorentzian Lie 3-algebras and in
this section we will continue by classifying the metric Lie 3-algebras
with signature $(2,p)$.  Clearly any such metric Lie 3-algebra will be
isomorphic to one of two types:
\begin{itemize}
\item $V_0 \oplus V_1 \oplus \dots$, where $V_0$ is an indecomposable
  metric Lie 3-algebra of signature $(2,*)$ and $V_{i\geq1}$ are
  indecomposable euclidean Lie 3-algebras; or
\item $V_1 \oplus V_2 \oplus V_3 \oplus \dots$, where $V_{1,2}$ are
  indecomposable lorentzian Lie 3-algebras and $V_{i\geq3}$ are
  indecomposable euclidean Lie 3-algebras.
\end{itemize}

The indecomposable euclidean Lie 3-algebras have been classified in
\cite{NagykLie} (see also \cite{GP3Lie,GG3Lie}) and the indecomposable
lorentzian Lie 3-algebras have been classified in \cite{Lor3Lie}.  It
remains to classify the indecomposable metric Lie 3-algebra of
signature $(2,*)$ and this is what the rest of this section is devoted
to.

\subsection{Notation}
\label{sec:notation}

In this section we will use the following notation.  Lie 3-algebras will
be denoted by capital letters $V,W,...$, whereas Lie algebras will be
denoted by lowercase fraktur letters $\fg,\fh,...$.

If $\fk$ is a metric Lie algebra, we will let $W(\fk)$ denote the metric
Lie 3-algebra $W(\fk) = \RR u \oplus \RR v \oplus \fk$ with
inner product which extends the ad-invariant inner product on $\fk$ by
declaring $u,v$ perpendicular to $\fk$ and, in addition,
$\left<u,u\right>=0=\left<v,v\right>$, $\left<u,v\right>=1$.  The Lie
3-brackets of $W(\fk)$ are given in terms of the inner product and the
Lie bracket of $\fk$ by
\begin{equation}
  \label{eq:Wk}
  [u,x,y] = [x,y] \qquad\text{and}\qquad
  [x,y,z] = - \left<[x,y],z\right> v~,
\end{equation}
for all $x,y,z\in\fk$.  This class of Lie 3-algebras was discovered
independently in \cite{GMRBL,BRGTV,HIM-M2toD2rev}.  We will mostly be
interested in the case where $\fk$ is a reductive Lie algebra with a
positive-definite inner product, so that $W(\fk)$ is lorentzian.  In
this case, if $\fk$ is not semisimple, then $W(\fk)$ will be
decomposable, since any abelian summands of $\fk$ will factorise.
Indeed, if we let $\fk = \fs \oplus \fa$ with $\fs$ semisimple and
$\fa$ abelian, then $W(\fs \oplus \fa) = W(\fs) \oplus A$, where $A$
is the abelian Lie 3-algebra sharing the same underlying vector space
as the abelian Lie algebra $\fa$.


\subsection{Structure of metric Lie 3-algebras with signature $(2,p)$}
\label{sec:structure-metric-lie}

Let $V$ be a finite-dimensional indecomposable metric Lie 3-algebra of
signature $(2,p)$.  The results of Section~\ref{sec:structure} allow
us to conclude that one of two situations can happen: either $V$ is
simple, whence isomorphic to $S_{2,2}$, or $V$ will be the double
extension by a one-dimensional Lie 3-algebra $U$ of a lorentzian Lie
3-algebra $W$.  In other words, $V = \RR(u,v) \oplus W$, where
$\left<u,u\right> = \left<v,v\right> = 0$ and $\left<u,v\right>=1$ and
$u,v \perp W$, and with 3-brackets
\begin{equation}
  \label{eq:V2p}
  [u,x,y] = [x,y] \qquad\text{and}\qquad
  [x,y,z] = [x,y,z]_W - \left<[x,y], z\right> v~,
\end{equation}
for all $x,y,z \in W$, which defines $[x,y]$ and where $[x,y,z]_W$ are
the 3-brackets of $W$.  The 3-Jacobi identity is equivalent to
$[-,-]:\Lambda^2 W \to W$ being a Lie (2-)bracket which leaves
invariant the Lie 3-bracket $[-,-,-]_W$.  Furthermore, both the Lie
algebra and the Lie 3-algebra structures on $W$ preserve the
lorentzian inner product.

$W$ is therefore simultaneously a lorentzian Lie algebra and a
lorentzian Lie 3-algebra, relative to the same inner product.  We shall
denote it $W$ as a Lie 3-algebra and as a vector space, but $\fw$ as a
Lie algebra.  As Lie 3-algebra we may write it as $W = W_0 \oplus W_1$,
where $W_0$ is an indecomposable lorentzian Lie 3-algebra and $W_1$ is a
euclidean Lie 3-algebra.  By the results of \cite{Lor3Lie}, $W_0$ can be
either one-dimensional, isomorphic to $S_{1,3}$ or else isomorphic to
$W(\fs)$, where $\fs$ is a euclidean semisimple Lie algebra, whereas by
the results of \cite{NagykLie} (see also \cite{GP3Lie,GG3Lie}), $W_1
\cong A \oplus S_{0,4} \oplus \dots \oplus S_{0,4}$, where $A$ is an
abelian euclidean Lie 3-algebra.  In summary, we have the following
possibilities for $W$, as a Lie 3-algebra:
\begin{itemize}
\item $W = A \oplus S_{0,4} \oplus \dots \oplus S_{0,4}$, where $A$ is a
  lorentzian abelian Lie 3-algebra;
\item $W = A \oplus S_{1,3} \oplus S_{0,4} \oplus \dots \oplus S_{0,4}$,
  where $A$ is a euclidean abelian Lie 3-algebra; and
\item $W = W(\fk) \oplus S_{0,4} \oplus \dots \oplus S_{0,4}$, where
  $\fk = \fs \oplus \fa$ is a euclidean reductive Lie algebra.
\end{itemize}

As a Lie algebra, the adjoint representation of $\fw$ on $W$ preserves
both the inner product and the 3-algebra structure.  It follows from
the results of \cite[§3.2]{Lor3Lie} that the adjoint representation of
$\fw$ preserves the subspaces corresponding to the simple factors and
hence also their perpendicular complement.  This means that each
summand of the vector space $W$ is a subrepresentation under the
adjoint representation and hence an ideal.  Furthermore, it also
follows from the results of \cite[§3.2]{Lor3Lie}, that any summands in
$W$ isomorphic to $S_{0,4}$ factor out, decomposing $V$ in the
process.  Hence for indecomposable $V$, they have to be absent.  This
means that the possibilities for $W$ as a Lie 3-algebra become
\begin{enumerate}
\renewcommand{\labelenumi}{(\Roman{enumi})}
\item $W = A$, where $A$ is a lorentzian abelian Lie 3-algebra;
\item $W = A \oplus S_{1,3}$, where $A$ is a euclidean abelian Lie 3-algebra; and
\item $W = W(\fs\oplus\fa)$;
\end{enumerate}
whereas for $W$ as a Lie algebra, we find that in the case (I), $\fw$ is
a lorentzian Lie algebra of dimension $\dim A$; whereas in case (II),
$\fw = \fg \oplus \fh_0$, where $\fg$ is a euclidean Lie algebra of
dimension $\dim A$ and $\fh_0$ is a four-dimensional lorentzian Lie
algebra, hence either abelian or isomorphic to one of the following
three Lie algebras: $\fso(3) \oplus \RR$, $\fso(1,2) \oplus \RR$ or the
solvable Nappi--Witten Lie algebra \cite{NW}, described as a central
extension of the Lie algebra of euclidean motions of $\RR^2$ or,
alternatively, as a double extension of an abelian two-dimensional
euclidean Lie algebra by a one-dimensional Lie algebra \cite{FSsug}.
Case (III) is somewhat different and will be treated in detail below.

Now we will show that case (II) always decomposes $V$ and hence it
cannot occur.  The proof is analogous to the one in \cite[§3.2]{Lor3Lie}
for the euclidean simple factors $S_{0,4}$.  We only need to show that
any of the nonabelian Lie algebras $\fh_0$ can occur as the reduced Lie
algebra associated to some $x \in S_{1,3}$; that is, the one with Lie
bracket $[x,-,-]$.  As mentioned above, there are three possible
nonabelian lorentzian four-dimensional Lie algebras:
\begin{itemize}
\item $\fso(3) \oplus \RR$, which has a timelike centre;
\item $\fso(1,2) \oplus \RR$, which has a spacelike centre; and
\item the solvable Nappi--Witten Lie algebra, which has a lightlike
  centre.
\end{itemize}
Given the causal characters of the centres, the following should perhaps
not be too surprising.

\begin{proposition}
  Let $0 \neq x\in S_{1,3}$ and let $\fh_x$ denote the Lie algebra
  structure on the vector space $S_{1,3}$ with Lie bracket $[x,-,-]$.
  Then
  \begin{itemize}
  \item if $x$ is timelike, $\fh_x \cong \fso(3) \oplus \RR$;
  \item if $x$ is spacelike, $\fh_x \cong \fso(1,2) \oplus \RR$; and
  \item if $x$ is lightlike, $\fh_x$ is isomorphic to the Nappi--Witten
    Lie algebra.
  \end{itemize}
\end{proposition}

\begin{proof}
  The 3-brackets of $S_{1,3}$, relative to a pseudo-orthonormal basis
  $(\be_0,\be_1,\be_2,\be_3)$, are given by
  \begin{equation*}
    [\be_0,\be_1,\be_2]=-\be_3 \qquad
    [\be_0,\be_1,\be_3]=+\be_2 \qquad
    [\be_0,\be_2,\be_3]=-\be_1 \qquad
    [\be_1,\be_2,\be_3]=-\be_0~.
  \end{equation*}
  The automorphism group of these brackets is $\SO(1,3)$, whence without
  loss of generality we can take $x$ to be $\be_0$, $\be_3$ or
  $\be_0+\be_3$ in the timelike, spacelike and lightlike cases,
  respectively.  We now discuss the reduced Lie algebras in each case.
  \begin{itemize}
  \item For $x = \be_0$, we have
    \begin{equation*}
    [\be_1,\be_2]=-\be_3 \qquad
    [\be_1,\be_3]=+\be_2 \qquad
    [\be_2,\be_3]=-\be_1~,
    \end{equation*}
    and in addition $\be_0$ central.  The resulting Lie algebra is
    clearly isomorphic to $\fso(3) \oplus \RR$.
  \item For $x = \be_3$, we have
    \begin{equation*}
    [\be_0,\be_1]=+\be_2 \qquad
    [\be_0,\be_2]=-\be_1 \qquad
    [\be_1,\be_2]=-\be_0~.
    \end{equation*}
    with $\be_3$ central.  The resulting Lie algebra is clearly
    isomorphic to $\fso(1,2) \oplus \RR$.
  \item Let $\be_\pm = \be_3 \pm \be_0$ and take $x = \be_+$, to obtain
    \begin{equation*}
      [\be_1,\be_2]=-\be_+ \qquad
      [\be_-,\be_1]=-2\be_2 \qquad
      [\be_-,\be_2]=+2\be_1~,
     \end{equation*}
     with $\be_+$ central.  We recognise this as the double extension of
     the the abelian Lie algebra spanned by $\be_1,\be_2$ by the
     one-dimensional Lie algebra spanned by $\be_-$ with dual $\be_+$.
     We can interpret $\be_1,\be_2$ as the generators of translations in
     the plane and $\be_-$ as the generator of rotations, and we are
     then centrally extending the translations by $\be_+$.  In either of
     these descriptions, we see that the resulting Lie algebra is
     isomorphic to the Nappi--Witten Lie algebra.
  \end{itemize}
\end{proof}

Essentially the same proof as that in \cite[§3.2]{Lor3Lie} for $S_{0,4}$
now shows that $S_{1,3}$ can be twisted out of $V$, decomposing it.  In
summary, case (2) cannot occur.

Finally, let us discuss case (III).  We will let $\fk = \fs \oplus \fa$
denote a generic reductive Lie algebra with a positive-definite
invariant inner product.  Under the adjoint representation, $\fw$ gets
mapped to the Lie algebra $\Der^0 W(\fk)$ of skewsymmetric derivations
of the Lie $3$-algebra $W(\fk)$, which we now determine.

\begin{proposition}\label{pr:Der0}
  Let $\fs$ and $\fa$ be a semisimple and abelian Lie algebras,
  respectively, with invariant positive-definite inner products.  Then
  \begin{equation*}
   \Der^0 W(\fs\oplus\fa)\cong \left(\ad\fs \ltimes
     \fs_{\text{ab}}\right) \oplus \left(\fso(\fa) \ltimes \fa\right)~,
  \end{equation*}
  with $\fs_{\text{ab}}$ and $\fa$ acting as null rotations on the
  lorentzian vector space $W(\fs \oplus \fa)$.
\end{proposition}

\begin{proof}
  The most general skewsymmetric endomorphism of $W(\fs\oplus\fa)$ is
  given in terms of $\alpha\in\RR$, $y,z\in\fs$, $b,c\in \fa$,
  $f\in\fso(\fs)$, $g\in\fso(\fa)$ and $\varphi: \fs \to \fa$ by
  \begin{equation*}
    \begin{aligned}[m]
      D \be_- &= \alpha \be_- + y + b\\
      D \be_+ &= -\alpha \be_+ + z + c
    \end{aligned}
    \qquad\qquad
    \begin{aligned}[m]
      D x &= -\left<z,x\right>\be_- - \left<y,x\right>\be_+ + f(x) +
      \varphi(x)\\
      D a &= -\left<c,a\right>\be_- - \left<b,a\right>\be_+ -
      \varphi^t(a) + g(a)~,
    \end{aligned}
  \end{equation*}
  for all $x\in\fs$ and $a\in \fa$.  Demanding that $D$ preserves the Lie
  3-bracket we obtain the following extra conditions:
  \begin{itemize}
  \item From $[\be_+,\be_-,x]$ we find $[z,x]_{\fs}=0$ for all $x$, but
    since $\fs$ is semisimple and has trivial centre, we conclude that $z=0$.
  \item From $[\be_-,x_1,x_2]$ we find that $\varphi$ must annihilate
    $[\fs,\fs]_{\fs}$, which implies that $\varphi = 0$ since $\fs$ is
    semisimple.  One also finds that $f + \alpha \id$ is a derivation of
    $\fs$ which, since all derivations are inner, allows us to conclude
    that $f + \alpha \id \in \ad\fs$.
  \item From $[a,x_1,x_2]$ we find that $\left<c,a\right>=0$ for all
    $a$, whence $c=0$.
  \item From $[x_1,x_2,x_3]$ we find that $f$ must be skewsymmetric,
    which means that $\alpha =0$, whence $f \in \ad\fs$.
  \end{itemize}
  In summary, the most general $D \in \Der^0W(\fs\oplus\fa)$
  is given in terms of $g \in \fso(\fa)$, $c\in \fa$ and $y,z\in\fs$ by
  \begin{equation*}
    \begin{aligned}[m]
      D \be_- &= z + c\\
      D \be_+ &= 0
    \end{aligned}
    \qquad\qquad
    \begin{aligned}[m]
      D x &= [y,x]_{\fs} - \left<z,x\right>\be_+\\
      D a &= g a- \left<c,a\right>\be_+~.
    \end{aligned}
  \end{equation*}
  The Lie bracket of $\Der^0W(\fs\oplus\fa)$ can be computed simply from
  the commutator of derivations, and we obtain
  \begin{equation*}
    \left[(y_1,z_1,g_1,c_1), (y_2,z_2,g_2,c_2)\right] =
    \left([y_1,y_2]_{\fs},[y_1,z_2]_{\fs}-[y_2,z_1]_{\fs},[g_1,g_2]_{\fso(\fa)},g_1c_2-g_2c_1\right)~,
  \end{equation*}
  which is precisely the direct sum of $\fs\ltimes \fs_{\text{ab}}$
  (with generic elements $(y,z)$) and $\fso(\fa)\ltimes \fa$ (with
  generic elements $(g,c)$).
\end{proof}

It follows that $W(\fs\oplus\fa)$ is not fully reducible as a
representation of $\Der^0W(\fs\oplus\fa)$.  Indeed, $\be_+$ spans an
invariant subspace without a complementary subspace which is also
invariant.

Proposition~\ref{pr:Der0} restricts the possible Lie algebra structures
$\fw$ on the underlying vector space of $W(\fs \oplus \fa)$, since for
every $w\in\fw$, $\ad_w \in \Der^0 W(\fs\oplus\fa)$.  In fact, we have
the following

\begin{proposition}\label{pr:preLieW}
  The most general Lie algebra $\fw$ is given by
  \begin{equation*}
    \begin{aligned}[m]
      [\be_-, a] &= J a\\
      [\be_-, s] &= [z,s]_{\fs}
    \end{aligned}
    \qquad\qquad
    \begin{aligned}[m]
      [s_1,s_2] &= [\psi s_1,s_2]_{\fs} + \left<z,[s_1,s_2]_{\fs}\right> \be_+\\
      [a_1,a_2] &= [a_1,a_2]_{\fr} + \left<J a_1,a_2\right> \be_+
    \end{aligned}
  \end{equation*}
  for all $a,a_1,a_2\in\fa$ and $s,s_1,s_2\in\fs$ and where $J \in
  \fso(\fa)$, $z\in\fs$, $[-,-]_{\fr}$ defines a
  reductive Lie algebra structure $\fr$ on the vector space $\fa$, $J$
  is a derivation over $[-,-]_{\fr}$, and $\psi \in \End\fs$ obeys
  \begin{equation}
    \label{eq:skewpsi}
    [\psi s_1,s_2]_{\fs} = [s_1,\psi s_2]_{\fs}
  \end{equation}
  and
  \begin{equation}
    \label{eq:psijacobi}
    \psi[\psi s_1,s_2]_{\fs} = [\psi s_1,\psi s_2]_{\fs}~,    
  \end{equation}
  for all $s_1,s_2\in\fs$, and
  \begin{equation}
    \label{eq:zpsi}
    [z,\psi s]_{\fs} = \psi [z,s]_{\fs}~,
  \end{equation}
  for all $s\in \fs$.
\end{proposition}

\begin{proof}
  In this proof, $W:=W(\fs\oplus\fa)$.  We determine the most general
  form of the Lie brackets using Proposition~\ref{pr:Der0} and the
  fact that $[x,y]$ can be interpreted both as $\ad_x y$ or as $-\ad_y
  x$, which are the actions of $\ad_x \in \Der^0W$ on $y$ and of
  $\ad_y \in \Der^0W$ on $x$, for all $x,y \in W$.

  Since every derivation $D \in \Der^0 W$ annihilates $\be_+$, we see
  that $\be_+$ is central in $\fw$.

  Now consider the bracket $[\be_-,a]$ for $a\in \fa$.  From
  Proposition~\ref{pr:Der0}, we see that $[\be_-,a] = \ad_{\be_-} a
  \in \fa \oplus \RR \be_+$, whereas $[\be_-, a] = \ad_{-a}\be_- \in
  \fa \oplus \fs$, whence $[\be_-, a] \in \fa$ and hence $\ad_{\be_-}
  : \fa \to \fa$ defines a skewsymmetric endomorphism we call $J$.
  Similarly, $[\be_-,s]$ for $s\in\fs$, belongs to $\fs$, whence
  $\ad_{\be_-} : \fs \to \fs$ defines a skewsymmetric endomorphism,
  which by Proposition~\ref{pr:Der0} is actually an inner derivation
  (relative to the Lie bracket of $\fs$), whence $[\be_-,s] =
  [z,s]_{\fs}$ for some $z\in \fs$.

  Now consider the bracket $[a,s]$ for $a\in\fa$ and $s \in \fs$.
  From Proposition~\ref{pr:Der0}, $\ad_a s  \in \fs \oplus \RR\be_+$,
  whereas $\ad_s a \in \fa \oplus \RR\be_+$, whence $[a,s] \in
  \RR\be_+$.  However, metricity shows that this is has to vanish,
  since $\left<[s,a],\be_-\right> = \left<s,[a,\be_-]\right> = 0$,
  since $[a,\be_-] \in \fa$ and $\fa \perp \fs$.

  Now consider the bracket $[s_1,s_2]$ for $s_i \in \fs$.  Since
  $[s_1,\be_-] = [s_1,z]$ and using metricity, we find that
  $[s_1,s_2] = [\psi s_1,s_2] + \left<z,[s_1,s_2]\right>\be_+$, for
  some endomorphism $\psi \in \End\fs$, \emph{not} necessarily a Lie
  algebra homomorphism.

  Finally, we consider the bracket $[a_1,a_2]$ for $a_i \in \fa$.
  Since $[a_1,\be_-]=-Ja_1$ and again using metricity, we find that
  $[a_1,a_2] = g(a_1)a_2 + \left<Ja_1,a_2\right>\be_+$, where $g: \fa
  \to \fso(\fa)$.

  There are some conditions that we have to impose on these brackets:
  skewsymmetry, Jacobi identity and that the map $\ad: \fw \to \Der^0
  W$ is a Lie algebra homomorphism.  These conditions are
  straight-forward to impose and can be summarised as follows.
  Skewsymmetry of the bracket and the Jacobi identity
  \begin{equation*}
    [a_1,[a_2,a_3]] = [[a_1,a_2],a_3] + [a_1,[a_2,a_3]]~.
  \end{equation*}
  imply that the bracket $[a_1,a_2]_{\fr} := g(a_1) a_2$ defines a
  reductive (since $\ad^{\fr} a \in \fso(\fa)$) Lie algebra structure,
  say $\fr$, on the vector space $\fa$.  The skewsymmetry of the Lie
  bracket on $\fw$ and the Jacobi identity
  \begin{equation*}
    [\be_-,[a_1,a_2]] = [[\be_-,a_1],a_2] + [a_1,[\be_-,a_2]]~,
  \end{equation*}
  say that $J \in \Der^0\fr$.

  Condition \eqref{eq:skewpsi} on $\psi$ follows from the skewsymmetry
  of the Lie bracket and \eqref{eq:zpsi} from the Jacobi identity
  \begin{equation*}
    [\be_-,[s_1,s_2]] = [[\be_-,s_1],s_2] + [s_1,[\be_-,s_2]]~.
  \end{equation*}
  Indeed, expanding the above Jacobi identity we see that
  \begin{equation*}
    [z,[\psi s_1,s_2]_{\fs}]_{\fs} = [\psi[z,s_1]_{\fs},s_2]_{\fs} +
    [\psi s_1,[z,s_2]_{\fs}]_{\fs}~,
  \end{equation*}
  which, upon using the Jacobi identity for $[-,-]_{\fs}$ on the
  left-hand side and \eqref{eq:skewpsi} on the first term of the
  right-hand side, becomes
  \begin{equation*}
    [[z,\psi s_1]_{\fs},s_2]_{\fs} = [[z,s_1]_{\fs},\psi s_2]_{\fs} =
    [\psi[z,s_1]_{\fs}, s_2]_{\fs}~.
  \end{equation*}
  Since $\fs$ is semisimple and has trivial centre, we see that this implies
  $[z,\psi s_1]_{\fs}  = \psi[z,s_1]_{\fs}$, which is \eqref{eq:zpsi}.
  Finally, the Jacobi identity
  \begin{equation*}
    [s_1,[s_2,s_3]] = [[s_1,s_2],s_3] + [s_1,[s_2,s_3]]
  \end{equation*}
  is equivalent to \eqref{eq:psijacobi}.  Indeed, the above Jacobi
  identity expands to
  \begin{equation*}
    [\psi s_1, [\psi s_2,s_3]_{\fs}]_{\fs} = [\psi [\psi s_1, s_2]_{\fs}, s_3]_{\fs} + [\psi
    s_2, [\psi s_1, s_3]_{\fs}]_{\fs}
  \end{equation*}
  up to central terms which vanish due to the Jacobi identity for
  $[-,-]_{\fs}$.  Using the Jacobi identity of $[-,-]_{\fs}$ on the
  above relation we find
  \begin{equation*}
    [[\psi s_1, \psi s_2]_{\fs}- \psi [\psi s_1, s_2]_{\fs},
    s_3]_{\fs} = 0~,
  \end{equation*}
  which, using that $\fs$ has trivial centre, becomes \eqref{eq:psijacobi}.
\end{proof}

\begin{remark}\label{re:impsi}
  Notice that \eqref{eq:psijacobi} says that the image of $\psi$ is a
  Lie subalgebra and that on its image, $\psi$ commutes with the
  restriction there of $\ad^{\fs}$.
\end{remark}

It turns out that \eqref{eq:skewpsi} implies \eqref{eq:psijacobi}.
This will be useful later because \eqref{eq:skewpsi} is linear in
$\psi$.

\begin{lemma}\label{le:psijacobi}
  Let $\fs$ be a metric Lie algebra and let $\psi\in\End\fs$ obey
  \eqref{eq:skewpsi} for all $s_1,s_2\in\fs$.  The new bracket
  $[s_1,s_2]:=[\psi s_1,s_2]_{\fs}$ obeys the Jacobi identity.
\end{lemma}

\begin{proof}
  The Jacobi identity of $[s_1,s_2]:=[\psi s_1,s_2]_{\fs}$ is
  equivalent to
  \begin{equation*}
    \Cyc_{1,2,3} [[s_1,s_2],s_3] = \Cyc_{1,2,3} [\psi [\psi
    s_1,s_2]_{\fs},s_3]_{\fs} = 0~,
  \end{equation*}
  for all $s_i \in \fs$, and where $\Cyc$ indicates cyclic
  permutations of the indices.  This is clearly equivalent to
  \begin{equation*}
    \Cyc_{1,2,3} \left<[[s_1,s_2],s_3],s_4\right> = 0~,
  \end{equation*}
  for all $s_i\in\fs$.  We now manipulate this expression using invariance of the inner
  product, equation \eqref{eq:skewpsi} and the Jacobi identity for $[-,-]_{\fs}$:
  \begin{align*}
    \Cyc_{1,2,3} \left<[\psi [\psi s_1,s_2]_{\fs},s_3]_{\fs},s_4\right> &= 
    \Cyc_{1,2,3} \left<[[\psi s_1,s_2]_{\fs}, \psi s_3]_{\fs},s_4\right>\\
    &= \Cyc_{1,2,3} \left<[\psi s_1,s_2]_{\fs}, [\psi s_3, s_4]_{\fs}\right>\\
    &= \Cyc_{1,2,3} \left<[\psi s_1,s_2]_{\fs}, [s_3, \psi s_4]_{\fs}\right>\\
    &= \Cyc_{1,2,3} \left<[[\psi s_1,s_2]_{\fs},s_3]_{\fs}, \psi s_4\right>\\
    &= \Cyc_{1,2,3} \left( \left<[\psi s_1,[s_2, s_3]_{\fs}]_{\fs}, \psi s_4\right> +
      \left<[[\psi s_1,s_3]_{\fs},s_2]_{\fs}, \psi s_4\right> \right)~.
  \end{align*}
  The last term on the RHS can be further rewritten as
  \begin{align*}
    \Cyc_{1,2,3} \left<[[\psi s_1,s_3]_{\fs},s_2]_{\fs}, \psi s_4\right> &=
    \Cyc_{1,2,3} \left<[[s_1, \psi s_3]_{\fs},s_2]_{\fs}, \psi s_4\right>\\
    &= \Cyc_{1,2,3} \left<[[s_2, \psi s_1]_{\fs},s_3]_{\fs}, \psi s_4\right>\\
    &= - \Cyc_{1,2,3} \left<[[\psi s_1,s_2]_{\fs},s_3]_{\fs}, \psi s_4\right>~,
  \end{align*}
  where we have used cyclicity and \eqref{eq:skewpsi}.  Therefore we see, using
  \eqref{eq:skewpsi} again and the invariance of the inner product, that
  \begin{align*}
    \Cyc_{1,2,3} \left<[\psi [\psi s_1,s_2]_{\fs},s_3]_{\fs},s_4\right> &= \half
    \Cyc_{1,2,3} \left<[\psi s_1,[s_2, s_3]_{\fs}]_{\fs}, \psi s_4\right> \\
    &= - \Cyc_{1,2,3} \left<[s_2, s_3]_{\fs}, [\psi s_1, \psi s_4]_{\fs}\right> \\
    &= \Cyc_{1,2,3} \left<[s_1,[s_2, s_3]_{\fs}]_{\fs}, \psi^2 s_4\right>~,
  \end{align*}
  which vanishes by the Jacobi identity of $[-,-]_{\fs}$.
\end{proof}

\begin{remark}
  In the lemma we assumed only that $\fs$ is a metric Lie algebra.
  However in order to relate the Jacobi identity of the bracket $[\psi
  s_1,s_2]_{\fs}$ to \eqref{eq:psijacobi} we did use, in addition, that
  $\fs$ had trivial centre.
\end{remark}

\begin{proposition}\label{pr:psisolved}
  Let $\fs = \fs_1 \oplus \dots \oplus \fs_p$ be a semisimple Lie
  algebra with $\fs_i$ its simple ideals, and let $\psi \in \End\fs$
  obey \eqref{eq:skewpsi} for all $s_1,s_2\in\fs$.  Then $\psi =
  \sum_{i=1}^p \lambda_i \Pi_i$, where $\Pi_i$ is the orthogonal
  projection onto $\fs_i$ and where $\lambda_i \in \RR$.
\end{proposition}

\begin{proof}
  We observe that if $\psi \in \End\fs$ obeys
  condition~\eqref{eq:skewpsi}, then so does any power $\psi^n$.
  Furthermore, condition~\eqref{eq:skewpsi} is clearly linear in
  $\psi$, whence any linear combination of endomorphisms satisfying
  condition~\eqref{eq:skewpsi} will again satisfy \eqref{eq:skewpsi}.
  Hence in particular the exponential $\exp(t \psi)$ satisfies
  \eqref{eq:skewpsi} and hence, by Lemma~\ref{le:psijacobi}, also
  equation \eqref{eq:psijacobi}.  By Remark \ref{re:impsi}, $\exp(t
  \psi)$ commutes with the adjoint representation on its image, which
  being invertible is all of $\fs$.  Since $\fs$ is semisimple,
  Schur's lemma says that $\exp(t \psi)$ is a scalar matrix on each
  simple factor; that is, $\exp(t\psi) = \sum_i \theta_i(t) \Pi_i$,
  with $\Pi_i$ the orthogonal projection onto $\fs_i$.
  Differentiating at $t=0$, we find $\psi = \sum_i \lambda_i \Pi_i$,
  where $\lambda_i = \theta_i'(0)$.
\end{proof}

In summary, we have proved the following refined version of Proposition
\ref{pr:preLieW}.

\begin{proposition}\label{pr:LieW}
  The most general compatible Lie algebra structure $\fw$ on the vector
  space $W(\fs\oplus\fa)$ is given by
  \begin{equation*}
    \begin{aligned}[m]
      [\be_-, a] &= J a\\
      [\be_-, s] &= [z,s]_{\fs}
    \end{aligned}
    \qquad\qquad
    \begin{aligned}[m]
      [s_1,s_2] &= [\psi s_1,s_2]_{\fs} + \left<z,[s_1,s_2]_{\fs}\right> \be_+\\
      [a_1,a_2] &= [a_1,a_2]_{\fr} + \left<J a_1,a_2\right> \be_+
    \end{aligned}
  \end{equation*}
  for all $a,a_1,a_2\in\fa$ and $s,s_1,s_2\in\fs$ and where $z\in\fs$,
  $[-,-]_{\fr}$ defines a reductive Lie algebra structure on $\fa$, $J
  \in \fso(\fa) \cap \Der\fr$, and $\psi \in \End\fs$ is given by $\psi
  = \sum_i \lambda_i \Pi_i$, where $\lambda_i \in \RR$ and $\Pi_i$ are
  the orthogonal projections onto the simple factors of $\fs$.
\end{proposition}

Imposing that the resulting metric Lie 3-algebra $V$ in \eqref{eq:V2p}
be indecomposable will further restrict the form of $\fw$, as will the
fact that $\fw$ is by construction a lorentzian Lie algebra and these
have been classified.  We turn to this now in order to finish the
classification of indecomposable metric Lie 3-algebras of signature
$(2,p)$.

\subsection{Indecomposable metric Lie 3-algebras of signature $(2,p)$}
\label{sec:indecomposable}

We saw above that there are two types of indecomposable Lie 3-algebras
$V$ of signature $(2,p)$, characterised by a lorentzian $p$-dimensional
vector space $W$ on which we have both a metric Lie 3-algebra structure,
also denoted $W$, and a metric Lie algebra structure, denoted $\fw$.
These two possible $W$s, said to be of types I and III above, are the
following:
\begin{itemize}
\item[(I)] $W = A$, abelian and $\fw$ any lorentzian Lie algebra; that
  is, $\ad\fw < \fso(\fw)$; and
\item[(III)] $W = W(\fs \oplus \fa)$ and $\fw$ given by
  Proposition~\ref{pr:LieW}, with some further restrictions to be
  explicited below.
\end{itemize}
We remark that although type~I is the special case of type~III with
$\fs=0$, it nevertheless pays to consider it separately.

\subsubsection{Type I}
\label{sec:type-I}

The indecomposable Lie 3-algebra with $W$ of type~I are such
that $V = \RR u \oplus \RR v \oplus A$, with $A$ an abelian lorentzian
Lie 3-algebra, with inner product which extends the one in $A$ by
declaring that $u,v \perp A$, $\left<u,u\right>=0=\left<v,v\right>$ and
$\left<u,v\right>=1$.  The Lie 3-brackets are
\begin{equation}
  \label{eq:type-I}
  [u,x,y] = [x,y] \qquad\text{and}\qquad
  [x,y,z] = - \left<[x,y],z\right> v~,
\end{equation}
with $[x,y]$ the Lie brackets of a lorentzian Lie algebra $\fw$, which
may be decomposable.  The most general lorentzian Lie algebra is given by
\begin{equation*}
  \fw = \fg_0 \oplus \ft \oplus \fs~,
\end{equation*}
where $\fg_0$ is an indecomposable lorentzian Lie algebra, $\ft$ is an
abelian euclidean Lie algebra and $\fs$ is a euclidean semisimple Lie
algebra.  It follows from \eqref{eq:type-I} that $\ft$ cannot appear, for
otherwise it decomposes $V$.  Hence, $\fw = \fg_0 \oplus \fs$.  The
indecomposable lorentzian Lie algebras $\fg_0$ have been classified.

\begin{theorem}[\cite{MedinaLorentzian}]\label{th:lorentzianLie}
  A finite-dimensional indecomposable lorentzian Lie algebra is either
  one-dimensional, isomorphic to $\fso(1,2)$ or else isomorphic to the
  solvable Lie algebra $\fm_J$, defined on the vector space $\RR
  \be_- \oplus \RR \be_+ \oplus E$ by the Lie bracket
  \begin{equation*}
    [\be_-,x] = J x \qquad\text{and}\qquad [x,y] = \left<J x, y\right>
    \be_+~,
  \end{equation*}
  for all $x,y\in E$ and where $\left<-,-\right>$ is a positive-definite
  inner product on $E$, $J \in \fso(E)$ is invertible, and we extend the
  inner product on $E$ to all of $V$ by declaring $\be_\pm \perp E$,
  $\left<\be_+,\be_-\right> =1$ and $\left<\be_\pm,\be_\pm\right> = 0$.
\end{theorem}

We cannot take $\fg_0$ to be one-dimensional, since this decomposes $V$,
whence $\fw = \fso(1,2) \oplus \fs$ or $\fw = \fm_J \oplus \fs$, which
we will call \textbf{type~Ia} and \textbf{type~Ib}, respectively.

In summary, (indecomposable) type~Ia metric Lie 3-algebras of signature
$(2,p)$ are constructed as follows.  The initial data consists of a
semisimple Lie algebra $\fs$ with a positive-definite ad-invariant inner
product (which is implicit in the notation) and a choice of invariant
inner product on $\fso(1,2)$, which comes down to a positive real number
which multiplies (the negative of) the Killing form.  The corresponding
indecomposable type~Ia metric Lie 3-algebra is denoted
$V_{\text{Ia}}(\fs)$.  The underlying vector space is $\RR(u,v) \oplus
\fso(1,2) \oplus \fs$ with $\left<u,u\right>=\left<v,v\right>=0$,
$\left<u,v\right>=1$, and all $\oplus$s orthogonal.  The nonzero Lie
3-brackets are given by
\begin{equation}
  \label{eq:type-Ia}
  \begin{aligned}[m]
    [u,x,y] &= [x,y]_{\fso(1,2)}\\
    [u,s_1,s_2] &= [s_1,s_2]_{\fs}
  \end{aligned}
  \qquad\qquad
  \begin{aligned}[m]
    [x,y,z] &= - \left<[x,y]_{\fso(1,2)},z\right> v\\
    [s_1,s_2,s_3] &= - \left<[s_1,s_2]_{\fs},s_3\right> v~,
  \end{aligned}
\end{equation}
for all $x,y,z\in \fso(1,2)$ and $s_i \in \fs$.

Similarly, (indecomposable) type~Ib metric Lie 3-algebras of signature
$(2,p)$ are constructed as follows.  The initial data consists of a
triple $(E,J,\fs)$ consisting of an even-dimensional euclidean space $E$
with a nondegenerate skewsymmetric endomorphism $J$, and a semisimple
Lie algebra $\fs$ with a positive-definite ad-invariant inner product
(which is implicit in the notation).  The corresponding indecomposable
type~Ib metric Lie 3-algebra is denoted $V_{\text{Ib}}(E,J,\fs)$.  The
underlying vector space is $\RR(u,v)\oplus \RR(\be_+,\be_-) \oplus E
\oplus \fs$ with $\left<u,u\right> = \left<v,v\right> =
\left<\be_\pm,\be_\pm\right> = 0$, $\left<u,v\right>=1$,
$\left<\be_+,\be_-\right>=1$ and all $\oplus$s orthogonal.  The nonzero
Lie 3-brackets are given by
\begin{equation}
  \label{eq:type-Ib}
  \begin{aligned}[m]
    [u,s_1,s_2] &= [s_1,s_2]_{\fs}\\
    [s_1,s_2,s_3] &= - \left<[s_1,s_2]_{\fs},s_3\right> v\\
  \end{aligned}
  \qquad\qquad
  \begin{aligned}[m]
    [u,\be_-,x] &= J x\\
    [u,x,y] &= \left<J x,y\right> \be_+\\
    [\be_-,x,y] &=  - \left<J x,y\right> v~,
  \end{aligned}
\end{equation}
for all $x,y\in E$, $s_i \in \fs$.

As discussed in §\ref{sec:no-ghosts}, a sufficient condition for the
decoupling of negative-norm states from the Bagger--Lambert
lagrangian, is that the Lie 3-algebra should admit a maximally
isotropic centre.  In signature $(2,p)$ this means a two-dimensional
isotropic centre.  This indeed happens in type~Ib,
since both $\be_+$ and $v$ are central and span an isotropic subspace.
Due to the simplicity of $\fso(2,1)$, this does not happen in type~Ia.

\subsubsection{Type III}
\label{sec:type-III}

In this case $W = W(\fs \oplus \fa)$ and $\fw$ is a lorentzian Lie
algebra structure on $W$ with brackets given by
Proposition~\ref{pr:LieW}.  Indecomposability forces the following
condition: the reductive Lie algebra structure $\fr$ on the vector space
$\fa$ must be semisimple, since any element in the centre of $\fr$ is
central in $V$ and of positive norm, whence it spans a nondegenerate
ideal.

At the same time, being lorentzian, the Lie algebra $\fw$ can
be one of the following:
\begin{itemize}
\item $\fw = \ft \oplus \fk$, where $\ft$ is lorentzian abelian and
  $\fk$ is euclidean semisimple;
\item $\fw = \fso(1,2) \oplus \ft \oplus \fk$, with $\ft$ and $\fk$
  euclidean abelian and semisimple, respectively; and
\item $\fw = \fm_J \oplus \ft \oplus \fk$, with $\fm_J$ defined in
  Theorem~\ref{th:lorentzianLie} and again $\ft$ and $\fk$
  euclidean abelian and semisimple, respectively.
\end{itemize}
Again we must exclude the ``middle third,'' this time because
$\ad\fso(1,2) \cong \fso(1,2)$ and there is no $\fso(1,2)$ subalgebra of
$\Der^0W(\fs\oplus\fa)$.  This leaves the first and the third cases,
which we will call \textbf{type IIIa} and \textbf{type IIIb},
respectively.

In type IIIa, we see that $\be_- \in \ft$ is central, whence in the
notation of Proposition~\ref{pr:LieW}, $z=0$ and $J=0$, whence we have
Lie algebra structures on $\fa$ and $\fs$ which leave invariant the
euclidean inner product.  This means that the corresponding Lie algebras
are reductive.  From Proposition~\ref{pr:psisolved}, it follows that the
Lie algebra structure on $\fs$ is isomorphic to an ideal of $\fs$.  We
already observed that the Lie algebra structure $\fr$ on $\fa$ has to be
semisimple, for any abelian factors would decompose $V$, assumed
indecomposable.  Therefore the data describing such a Lie 3-algebra is
$(\fs,\fr,\lambda_1,\dots,\lambda_p)$, where $\fs= \fs_1\oplus \dots
\fs_p$ is a semisimple Lie algebra decomposed into its simple ideals,
$\lambda_i\in\RR$ and $\fr$ is a semisimple Lie algebra.  Both $\fs$ and
$\fr$ are equipped with positive-definite invariant inner products.
We let $\psi=\sum_{i=1}^p \lambda_i \Pi_i$, with $\Pi_i: \fs \to \fs_i$
denoting the orthogonal projection on $\fs_i$.  The corresponding
indecomposable Lie 3-algebra is denoted
$V_{\text{IIIa}}(\fs,\fk,\lambda_i)$.  The underlying vector space is
$\RR(u,v) \oplus \RR(\be_+,\be_-) \oplus \fs \oplus \fr$ with 
$\left<u,u\right> = \left<v,v\right> =
\left<\be_\pm,\be_\pm\right> = 0$, $\left<u,v\right>=1$,
$\left<\be_+,\be_-\right>=1$ and all $\oplus$s orthogonal.  The nonzero
Lie 3-brackets are given by
\begin{equation}
  \label{eq:type-IIIa}
  \begin{aligned}[m]
    [u, s_1,s_2] &= [\psi s_1,s_2]_{\fs}\\
    [u, a_1,a_2] &= [a_1,a_2]_{\fr}\\
    [\be_-,s_1,s_2] &= [s_1,s_2]_{\fs}
  \end{aligned}
  \qquad\qquad
  \begin{aligned}[m]
    [s_1,s_2,s_3] &= - \left<[s_1,s_2]_{\fs},s_3\right>\be_+ -
    \left<[\psi s_1,s_2]_{\fs},s_3\right> v\\
    [a_1,a_2,a_3] &= - \left<[a_1,a_2]_{\fr},a_3\right> v~.\\
  \end{aligned}
\end{equation}

Finally, let us consider case IIIb.  In this case, $\fw \cong \fm_J
\oplus \ft \oplus \fk$, where $\fk$ is semisimple, $\ft$ is abelian, and
$\fm_J$ is one of the indecomposable lorentzian Lie algebras in
Theorem~\ref{th:lorentzianLie}.  Consider the Lie algebra $\fw$ in
Proposition~\ref{pr:LieW}.  It will be convenient to split $\fs$ and
$\fa$ further as follows.  Let $\fs = \fg \oplus \fh$, where $\fh = \ker
\psi$ and $\fg = \im \psi$.  This split corresponds to partitioning
the simple ideals $\fs_i$ of $\fs$ into two sets, depending on whether
$\lambda_i$ is or is not zero.  As a result both $\fg$ and $\fh$ are
semisimple and commute with each other.  Similarly let us decompose the
reductive Lie algebra $\fr$ (with underlying vector space $\fa$) into
$\fr = \fl \oplus \fb$, where $\fl$ is semisimple and $\fb$ is abelian.
Relative to such splits, the Lie bracket of $\fw$ becomes
\begin{equation*}
  \begin{aligned}[m]
    [\be_-,s] &= [z,s]_{\fs}\\
    [\be_-,a] &= J a\\
    [g_1,g_2] &= [\psi g_1,g_2]_{\fg} + \left<z,[g_1,g_2]_{\fg}\right>\be_+
  \end{aligned}\quad\qquad
  \begin{aligned}[m]
    [h_1,h_2] &= \left<z,[h_1,h_2]_{\fh}\right>\be_+\\
    [\ell_1,\ell_2] &= [\ell_1,\ell_2]_{\fl} + \left<J \ell_1, \ell_2\right>\be_+\\
    [b_1,b_2] &= \left<J b_1, b_2\right>\be_+~,
  \end{aligned}
\end{equation*}
for all $a\in\fa$, $s\in\fs$, $g_i\in\fg$, $h_i\in\fh$, $\ell_i\in\fl$ and
$b_i\in\fb$.

The first observation is that because $\fl$ and $\fg$ are semisimple,
they do not admit nontrivial central extensions.  This means that we can
eliminate the component of $J$ in $\End\fl$ and the component of $z$ in
$\fg$ via an isometry, as we will now see.  But first a preliminary
result.

\begin{lemma}
  Under the decomposition $\fr = \fl \oplus \fb$, with $\fl$ semisimple
  and $\fb$ abelian, $J \in \Der^0\fr$ decomposes as
  \begin{equation*}
    J =
    \begin{pmatrix}
      J_{\fl} & 0 \\
      0 & J_{\fb}
    \end{pmatrix}~,
  \end{equation*}
  where $J_{\fl} \in \Der^0\fl$ and $J_{\fb} \in \fso(\fb)$.
\end{lemma}

\begin{proof}
  Since $J$ is a derivation, $J[\ell_1,\ell_2] = [J\ell_1,\ell_2] +
  [\ell_1,J\ell_2]$, which, since $\fb$ commutes with $\fl$, shows that
  $J[\ell_1,\ell_2] \in \fl$.  Since $\fl$ is semisimple, this shows
  that $J\ell \in \fl$ for all $\ell \in \fl$.  Since $\fl$ and $\fb$
  are orthogonal, we have for all $\ell\in\fl$ and $b\in\fb$, $0 =
  \left<J \ell,b\right> = - \left<\ell, Jb\right>$, whence $Jb \in
  \fb$.  Defining $J_{\fl}$ and $J_{\fb}$ the restrictions of $J$ to
  $\fl$ and $\fb$, respectively, we see that $J$ is as shown.
\end{proof}

Let $J_{\fl} \in \Der^0\fl$ denote the restriction of $J$ to $\fl$.
Since $\ell$ is semisimple, $J_{\fl}(\ell) = [\ell_0,\ell]$ for some
$\ell_0\in\fl$.  Let $f_1: \fw\to\fw$ be the isometry which sends $\ell
\mapsto \ell + \left<\ell_0,\ell\right> \be_+$ and $\be_- \mapsto \be_-
- \ell_0 - \half |\ell_0|^2 \be_+$ and is the identity elsewhere.  Then
$J_{\fl}$ does not appear in the Lie brackets of $f_1(\fw)$.  Now let
$f_2: \fw\to\fw$ denote the isometry which sends $g \mapsto g +
\left<\psi^{-1}z,g\right>\be_+$ and $\be_- \mapsto \be_- - \psi^{-1}
z_{\fg} - \half |\psi^{-1}z_{\fg}|^2\be_+$ (and is the identity
elsewhere), where $z_{\fg}$ is the projection of $z$ onto $\fg$ along
$\fh$.  Then $z_{\fg}$ does not appear in the Lie brackets of
$f_2(f_1(\fw))$.

We may therefore take $z \in \fh$ and $J\in\End\fb$ without loss of
generality.  The resulting nonzero Lie brackets for $\fw$ are then
\begin{equation*}
  \begin{aligned}[m]
    [\be_-,h] &= [z,h]_{\fh}\\
    [\be_-,b] &= J b\\
    [g_1,g_2] &= [\psi g_1,g_2]_{\fg}
  \end{aligned}\qquad\qquad
  \begin{aligned}[m]
    [h_1,h_2] &= \left<z,[h_1,h_2]_{\fh}\right>\be_+\\
    [\ell_1,\ell_2] &= [\ell_1,\ell_2]_{\fl}\\
    [b_1,b_2] &= \left<J b_1, b_2\right>\be_+~,
  \end{aligned}
\end{equation*}
for all $g_i\in\fg$, $h,h_i\in\fh$, $\ell_i\in\fl$ and $b,b_i\in\fb$, and
where $J \in \fso(\fb)$ and $z \in \fh$.  It follows from these brackets
that if $b \in \ker J \cap \fb$, then it is central in $V$ and has
positive norm, whence $V$ becomes decomposable.  Therefore for
indecomposability of $V$ we require $J$ to be nondegenerate when
restricted to $\fb$, whence $\fb$ must be even-dimensional.

As we now show, we can put $z=0$ without loss of generality.  The
following 3-brackets are the only ones in the type IIIb Lie 3-algebra
$V$ which involve $z$ and $\fh$:
\begin{equation*}
  \begin{aligned}[m]
    [u,\be_-,h] &= [z,h]_{\fh}\\
    [u,h_1,h_2] &= \left<z,[h_1,h_2]_{\fh}\right>\be_+
  \end{aligned}\qquad\qquad
  \begin{aligned}[m]
    [\be_-,h_1,h_2] &= [h_1,h_2]_{\fh} - \left<z,[h_1,h_2]_{\fh}\right> v\\
    [h_1,h_2,h_3] &= -\left<[h_1,h_2]_{\fh},h_3\right> \be_+~,
  \end{aligned}
\end{equation*}
where $h_i \in \fh$.  Consider the isometry $f : V \to V$ mapping $h
\mapsto h - \left<z,h\right>v$ and $u \mapsto u + z - \half |z|^2 v$ and
equal to the identity elsewhere.  The induced brackets in $f(V)$ are
formally the same, but with $z=0$.  We will therefore put $z=0$ from now
on.

Doing so in the Lie brackets for $\fw$ we obtain
\begin{equation*}
  \begin{aligned}[m]
    [\be_-,b] &= J b\\
    [b_1,b_2] &= \left<J b_1, b_2\right>\be_+
  \end{aligned}\qquad\qquad
  \begin{aligned}[m]
    [g_1,g_2] &= [\psi g_1,g_2]_{\fg}\\
    [\ell_1,\ell_2] &= [\ell_1,\ell_2]_{\fl}~,
  \end{aligned}
\end{equation*}
As expected, the resulting Lie algebra $\fw$ is isomorphic to $\fm_J
\oplus (\fg \oplus \fl) \oplus \fh$, where $\fg \oplus \fl$ is
semisimple, $\fh$ is abelian, and where the euclidean space $E$ in
$\fm_J$ is $\fb$.

It follows that the data defining a type IIIb indecomposable metric Lie
3-algebra of signature $(2,p)$ is the following: three semisimple Lie
algebras $\fg$, $\fh$ and $\fl$ each with a choice of euclidean inner
product, an \emph{invertible} endomorphism $\psi= \sum_i \lambda_i
\Pi_i$ of $\fg$, where $0 \neq \lambda_i \in \RR$ and $\Pi_i$ are the
orthogonal projections onto the simple ideals of $\fg$, and an
even-dimensional euclidean vector space $E$ with a nondegenerate
$J\in\fso(E)$.  The resulting type IIIb Lie 3-algebra, denoted
$V_{\text{IIIb}}(E,J,\fl,\fh,\fg,\psi)$, has as underlying
vector space $\RR(u,v) \oplus \RR(\be_+,\be_-) \oplus E \oplus\fl
\oplus\fh\oplus \fg$ with $\left<u,u\right> = \left<v,v\right> =
\left<\be_\pm,\be_\pm\right> = 0$,
$\left<u,v\right>=1=\left<\be_+,\be_-\right>$ and all $\oplus$s
orthogonal.  The nonzero Lie 3-brackets are given by
\begin{equation}
  \label{eq:type-IIIb}
  \begin{aligned}[m]
    [u,\be_-,x] &= J x\\
    [u,x,y] &= \left<J x,y\right>\be_+\\
    [\be_-,x,y] &= - \left<Jx,y\right> v\\
    [\be_-,h_1,h_2] &= [h_1,h_2]_{\fh}\\
    [h_1,h_2,h_3] &= -\left<[h_1,h_2]_{\fh},h_3\right> \be_+
  \end{aligned}\qquad
  \begin{aligned}[m]
    [u,g_1,g_2] &= [\psi g_1,g_2]_{\fg}\\
    [\be_-,g_1,g_2] &= [g_1,g_2]_{\fg}\\
    [g_1,g_2,g_3] &= - \left<[g_1,g_2]_{\fg},g_3\right> \be_+ - \left<[\psi g_1,g_2]_{\fg}, g_3\right> v\\
    [u,\ell_1,\ell_2] &= [\ell_1,\ell_2]_{\fl}\\
    [\ell_1,\ell_2,\ell_3] &= -\left<[\ell_1,\ell_2]_{\fl},\ell_3\right> v~,
  \end{aligned}
\end{equation}
where $x,y\in E$, $h,h_i\in\fh$, $g_i\in\fg$ and $\ell_i\in\fl$.

We can recognise several subalgebras among the above 3-brackets.
First of all we recognise two decomposable subalgebras: one isomorphic
to $W(\fl \oplus \RR^{1,1})$, where the $\RR^{1,1}$ is the abelian two-dimensional
lorentzian Lie algebra spanned by $\be_\pm$, and one isomorphic to
$W(\fh \oplus \RR^{1,1})$, with $u,v$ spanning the $\RR^{1,1}$.  Then we
have $V(E,J)$, defined by
\begin{equation}
  \label{eq:VEJ}
    [u,\be_-,x] = J x \qquad\qquad
    [u,x,y] = \left<J x,y\right>\be_+ \qquad\qquad
    [\be_-,x,y] = - \left<Jx,y\right> v~,
\end{equation}
for all $x,y \in E$.  It is an indecomposable metric Lie 3-algebra with
signature $(2,*)$.  We then have $V(\fg,\psi)$, with 3-brackets
\begin{equation}
  \label{eq:Vgpsi}
  \begin{aligned}[m]
    [u,g_1,g_2] &= [\psi g_1,g_2]_{\fg}\\
    [\be_-,g_1,g_2] &= [g_1,g_2]_{\fg}\\
    [g_1,g_2,g_3] &= - \left<[g_1,g_2]_{\fg},g_3\right> \be_+ - \left<[\psi g_1,g_2]_{\fg}, g_3\right> v~.
  \end{aligned}
\end{equation}
If $\psi = \lambda \id$ is a scalar endomorphism, $V(\fg,\psi)$ is
decomposable.  Indeed, in this case define
\begin{equation*}
  \hat u = \half (\lambda^{-1}u + \be_-) \qquad
  \hat v = \lambda v  + \be_+ \qquad
  \hat\be_+ = \half (\lambda v - \be_+) \qquad
  \hat \be_- = \lambda^{-1} u - \be_-~.
\end{equation*}
Then $[\hat\be_\pm,\fg,\fg]=0$, whence $V(\fg,\psi) \cong W(\fg \oplus
\RR^{1,1})$.  Furthermore, if $\fh$ and $\fl$ are absent, then if $\psi$
is again a scalar a short calculation shows that
$V_{\text{IIIb}}(E,J,\fzero,\fzero,\fg,\lambda \id) \cong
V_{\text{IIIb}}(E,-\lambda^{-1} J,\fzero,\fg,\fzero,0)$.

Finally let us remark that type Ib is the special case of type IIIb
where $\fh$ and $(\fg,\psi)$ are absent; and, similarly, type
IIIa is the special case of IIIb where $\fh$ and $(E,J)$ are absent.
In summary, we have proved the following

\begin{theorem}\label{th:2p}
  Let $V$ be an indecomposable metric Lie 3-algebra with signature
  $(2,p)$.  Then it is isomorphic to either $V_{\text{Ia}}(\fs)$ or
  $V_{\text{IIIb}}(E,J,\fl,\fh,\fg,\psi)$, which have been
  defined in \eqref{eq:type-Ia} and \eqref{eq:type-IIIb}, respectively.
  If the centre of $V$ contains a maximally isotropic plane, then it is
  of type IIIb, otherwise it is of type Ia.
\end{theorem}

The type IIIb algebras actually encapsulate a large class of metric Lie
3-algebras depending on which of the data $(E,J)$, $\fk$, $\fl$ or
$(\fg,\psi)$ is present.  It is worth listing the possible cases which
can occur, because they do tend to have different properties and, as
explained in Section~\ref{sec:3-BL}, many physically desirable
properties of the Bagger--Lambert model can be translated into
properties of the Lie 3-algebra.  Ignoring the trivial case, where none
of the structures are present, we have a priori 15 different types of
metric Lie 3-algebras.  However some of these types are isomorphic and
need not be counted as different.  To see this, we notice that $V(E,J)$
admits an automorphism $(u,v,\be_-,\be_+) \mapsto (-\be_-, -\be_+, u,
v)$ (and the identity on $E$) which preserves both the 3-brackets and
the inner product.  Under this map, the subalgebras $W(\fh) \oplus
\RR^{1,1}$ and $W(\fl) \oplus \RR^{1,1}$ of
$V_{\text{IIIb}}(E,J,\fh,\fl,\fg, \psi)$ get mapped to each other, and
$V(\fg,\psi)$ is mapped to $V(-\psi\fg, -\psi^{-1})$, where $-\psi\fg$
means the Lie algebra with bracket $[x,y]_{-\psi\fg} := [-\psi
x,y]_{\fg}$.  This means that type IIIb splits into 11 different types:
\begin{enumerate}
\item $V(E,J)$;
\item $W(\fh)\oplus\RR^{1,1}$, which is in the same class as
  $W(\fl)\oplus\RR^{1,1}$ and $V(\fg,\psi)$ with $\psi=\lambda\id$ a scalar;
\item $V(\fg,\psi)$, $\psi$ not a scalar;
\item $V_{\text{IIIb}}(E,J,\fzero,\fzero,\fg,\psi)$, $\psi$ again not a
  scalar;
\item $V_{\text{IIIb}}(E,J,\fh,\fzero,\fzero,0)$, which is
  in the same class as   $V_{\text{IIIb}}(E,J,\fzero,\fzero,\fg,\psi)$
  for $\psi$ a scalar and $V_{\text{IIIb}}(E,J,\fzero,\fl,\fzero,0)$;
\item $V_{\text{IIIb}}(0,0,\fh,\fl,\fzero,0)$;
\item $V_{\text{IIIb}}(0,0,\fh,\fzero,\fg,\psi)$, which is in
  the same class as $V_{\text{IIIb}}(0,0,\fzero,\fl,\fg,\psi)$;
\item $V_{\text{IIIb}}(E,J,\fh,\fl,\fzero,0)$;
\item $V_{\text{IIIb}}(0,0,\fh,\fl,\fg, \psi)$;
\item $V_{\text{IIIb}}(E,J,\fh,\fzero,\fg, \psi)$, which is in
  the same class as $V_{\text{IIIb}}(E,J,\fzero,\fl,\fg, \psi)$; and
\item the full $V_{\text{IIIb}}(E,J,\fh,\fl,\fg, \psi)$.
\end{enumerate}
We notice that all but the second and sixth cases are indecomposable.

\section{The Lie algebra of derivations}
\label{sec:derivations}

In this section we will consider the Lie algebras of derivations of the
metric Lie 3-algebras classified in Section~\ref{sec:indecomposable}.
In particular we will deconstruct the type IIIb Lie 3-algebra found in
the previous section and discuss some natural special cases.  Let us
make some generic remarks about automorphisms and derivations.

Given a metric Lie 3-algebra $V$ there are several groups of
automorphisms which are of interest.  The largest of such groups is
the group $\Aut V$ consisting of all automorphisms of $V$:
\begin{equation}
  \label{eq:AutV}
  \Aut V = \left\{\varphi \in \GL(V) \middle | \varphi[x,y,z]=[\varphi
    x, \varphi y, \varphi z],~\forall x,y,z\in V\right\}~.
\end{equation}
Because $V$ possesses an inner product, it is natural to restrict
ourselves to the subgroup $\Aut^c V$ of $\Aut V$ consisting of
automorphisms which rescale the inner product:
\begin{equation}
  \label{eq:AutcV}
  \Aut^c V = \left\{\varphi \in \Aut V \middle | \left<\varphi
      x,\varphi y\right> = \mu \left<x,y\right>,~\forall x,y\in
    V,~\exists \mu \in \RR\right\}~,
\end{equation}
which will be denoted \textbf{conformal automorphisms}.  Similarly, we
can restrict to automorphisms which preserve the inner product, namely
the \textbf{orthogonal automorphisms}
\begin{equation}
  \label{eq:Aut0V}
  \Aut^0 V = \left\{\varphi \in \Aut V \middle | \left<\varphi
      x,\varphi y\right> = \left<x,y\right>,~\forall x,y\in
    V\right\}~.
\end{equation}
Finally, we have the so-called \textbf{inner automorphisms}, which are
obtained by exponentiating the inner derivations.  We will call this
group $\Ad V$.  It is clear that we have the following chain of
inclusions
\begin{equation*}
  \Ad V < \Aut^0 V < \Aut^c V < \Aut V~.
\end{equation*}
Their Lie algebras are, respectively, the Lie algebras of derivations,
conformal derivations, skewsymmetric derivations and inner
derivations, giving rise to a similar chain of inclusions
\begin{equation*}
  \ad V \lhd \Der^0 V < \Der^c V < \Der V~,
\end{equation*}
with $\ad V$ the ideal of inner derivations, which generate the gauge
transformations in the Bagger--Lambert theory.  This allows us to
think of $\Ad V$ as the gauge group of the theory.

\subsection{Type Ia}
\label{sec:aut-type-Ia}

The 3-brackets of $V_{\text{Ia}}(\fs)$ are given in
\eqref{eq:type-Ia}, making $V_{\text{Ia}}(\fs) \cong W(\fso(1,2)
\oplus \fs)$, whose automorphisms were determined in
\cite[Proposition~11]{Lor3Lie}, which we recall here for convenience.

\begin{proposition}\label{pr:AutVIa}
  Every 3-algebra automorphism $\varphi \in \Aut V_{\text{Ia}}(\fs)$ is given by
  \begin{align*}
    \varphi(v) &= \beta^{-3} v\\
    \varphi(u) &= \beta u + \gamma v + t\\
    \varphi(x) &= \beta^{-1} a(x) - \beta^{-2} \left<t,a(x)\right> v~,
  \end{align*}
  for all $x\in\fso(1,2)\oplus\fs$ and where $\beta\in\RR^\times$,
  $\gamma\in\RR$, $t\in\fso(1,2)\oplus\fs$ and $a\in\SO(1,2) \times
  \Aut^0\fs$, where $\Aut^0\fs$ denotes the subgroup of automorphisms
  of $\fs$ which preserve the inner product.
\end{proposition}

Restricting to those automorphisms which preserve the inner product up
to a homothety, we find that $\gamma$ is fixed in terms of $\beta$ and
the norm of $t$:

\begin{proposition}\label{pr:AutcVIa}
  Every 3-algebra conformal automorphism $\varphi \in \Aut^c
  V_{\text{Ia}}(\fs)$ is given by
  \begin{align*}
    \varphi(v) &= \beta^{-3} v\\
    \varphi(u) &= \beta u -\half \beta |t|^2 v + t\\
    \varphi(x) &= \beta^{-1} a(x) - \beta^{-2} \left<t,a(x)\right> v~,
  \end{align*}
  for all $x\in\fso(1,2)\oplus\fs$ and where $t\in\fso(1,2)\oplus\fs$
  and $a\in\SO(1,2) \times \Aut^0\fs$.  The inner product is rescaled
  by $\beta^{-2}$: $\left<\varphi x, \varphi y\right> =
  \beta^{-2}\left<x,y\right>$.
\end{proposition}

Restricting to automorphisms which preserve the inner product we find
\cite[Proposition~12]{Lor3Lie} that $\beta=1$.

\begin{proposition}\label{pr:Aut0VIa}
  Every 3-algebra automorphism $\varphi \in \Aut^0 V_{\text{Ia}}(\fs)$
  preserving the inner product is given by
  \begin{equation*}
    \varphi(v) = v\qquad
    \varphi(u) = u -\half |t|^2 v + t\qquad
    \varphi(x) = a(x) - \left<t,a(x)\right> v~,
  \end{equation*}
  for all $x\in\fso(1,2)\oplus\fs$ and where $t\in\fso(1,2)\oplus\fs$
  and $a\in\SO(1,2) \times \Aut^0\fs$.
\end{proposition}

As shown in \cite{Lor3Lie}, the connected component of
$\Aut^0V_{\text{Ia}}(\fs)$ is $\Ad V_{\text{Ia}}(\fs)$, consisting of
the inner automorphisms obtained by exponentiating the inner
derivations of the Lie 3-algebra $V_{\text{Ia}}(\fs)$.  This gives the
following.

\begin{proposition}\label{pr:Der0VIa}
  Let $\Der^0V_{\text{Ia}}(\fs)$ denote the Lie algebra of skewsymmetric
  derivations of the Lie 3-algebra $V_{\text{Ia}}(\fs)$.  Then
  \begin{equation*}
    \Der^0V_{\text{Ia}}(\fs) \cong \left(\fso(1,2) \ltimes \RR^3 \right)
    \oplus \left(\fs \ltimes \fs_{\text{ab}}\right) \cong \ad
    V_{\text{Ia}}(\fs)~.
  \end{equation*}
\end{proposition}

The Lie algebra $\Der V_{\text{Ia}}(\fs)$ of $\Aut V_{\text{Ia}}(\fs)$
consists of derivations of $V$.  It is isomorphic to the real Lie
algebra with generators $D$, $S$, $L_x$ and $T_x$ for
$x\in\fso(1,2)\oplus\fs$, subject to the following nonzero Lie
brackets: 
\begin{equation*}
  [D,S] = -4 S~, \qquad
  [D,T_x] = - 2 T_x~, \qquad
  [L_x,L_y] = L_{[x,y]} \qquad\text{and}\qquad
  [L_x,T_y] = T_{[x,y]}~.
\end{equation*}
If we let $\fa$ denote the two-dimensional solvable Lie subalgebra
spanned by $D$ and $S$, then we find that $\Der V_{\text{Ia}}(\fs)$
has the following structure
\begin{equation*}
  \Der V_{\text{Ia}}(\fs) \cong \fa \ltimes \ad V_{\text{Ia}}(\fs)~.
\end{equation*}

\subsection{Type IIIb}
\label{sec:aut-type-IIIb}

Finally let us consider $V:=V_{\text{IIIb}}(E,J,\fl,\fh,\fg,\psi)$, a
general type IIIb Lie 3-algebra as defined in \eqref{eq:type-IIIb}.  As
mentioned at the end of Section~\ref{sec:type-III}, this type consists
of 9 different types of indecomposable Lie 3-algebras, depending on
which of the four ingredients $(E,J)$, $\fl$, $\fh$ or $(\fg,\psi)$ are
present.

Our strategy will be the following.  We will first write down the most
general endomorphism of $V$ taking into account that derivations
preserve the centre and the first derived ideal $V'=[V V V]$.  In
particular, elements that are not in $V'$ can not appear in the image of
elements of $V'$.  Then we will impose the derivation property to derive
constraints which do not depend on which of the ingredients are present.
Finally we will consider those constraints which depend on the presence
of a particular ingredient.  We will omit the routine details and simply
list the results.

For all type IIIb algebras, $\be_+$, $v$ are central, whereas in some
cases $\be_-$ or $u$ are central as well.  Those cases, however, are
decomposable and we shall ignore them in this section.  Similarly we
observe that $\be_-$ or $u$ do not belong to the first derived ideal.
The most general $D \in \End V$ that preserves the the centre and the
first derived ideal is given by
\begin{equation}\label{eq:EndV}
  \begin{aligned}[m]
    D\be_+ &= \alpha \be_+ + \beta v \\
    D v &= \gamma \be_+ + \delta v \\
    D\be_- &= a \be_- + b u + x_- + h_- + \ell_- + g_- + \eta \be_+ + \xi v \\
    D u &= c \be_- + d u + x_u + h_u + \ell_u + g_u + \theta \be_+ + \omega v\\
    Dx &= \varphi_{EE}(x) + \varphi_{E\fh}(x) + \varphi_{E\fl}(x) + \varphi_{E\fg}(x) + A_{1}(x) \be_+ + C_{1}(x) v \\
    Dh &= \varphi_{\fh E}(h) + \varphi_{\fh\fh}(h) + \varphi_{\fh \fl}(h) + \varphi_{\fh \fg}(h) + A_{2}(h) \be_+ + C_{2}(h) v \\
    D\ell &= \varphi_{\fl E}(\ell) + \varphi_{ \fl \fh}(\ell) + \varphi_{\fl \fl}(\ell) + \varphi_{\fl \fg}(\ell) + A_{3}(\ell) \be_+ + C_{3}(\ell) v \\
    Dg &= \varphi_{\fg E}(g) + \varphi_{\fg \fh}(g) + \varphi_{\fg \fl}(g) + \varphi_{\fg \fg}(g) + A_{4}(g) \be_+ + C_{4}(g) v ~,
  \end{aligned}
\end{equation}
where $\alpha, \beta, \gamma, \delta, a, b, c, d, \eta, \xi, \theta,
\omega \in \RR$, $A_i, C_i$ are in $E^*$, $\fh^*$, $\fl^*$ and $\fg^*$,
respectively for $i = 1,\dots,4$, $x_-, x_u \in E$, $h_-, h_u \in \fh$,
$\ell_-, \ell_u \in \fl$, $g_-, g_u \in \fg$ and $\varphi_{V_1 V_2}: V_1
\to V_2$ are linear maps.

Notice that from the fact that elements on the different subspaces $\fh,
\fl, \fg$ and $E$ never appear together on a 3-bracket and $u$, $e_-$
only appear together with elements in $E$, we find that if $V_1 \neq
V_2$ then $\varphi_{V_1 V_2} = 0$.  Also, from the vanishing brackets
$[u, h_1, h_2] = 0$ and $[e_-, \ell_1, \ell_2] = 0$ we find $\ell_- =
h_u = 0$.

We apply now the derivation $D$ to all other brackets and obtain the
following general map.  It is implicit that if $V$ does not include $E$,
then $x_- = x_u = 0$ and $Dx$ does not appear.  Similarly, if $\fh$ was
not there, then $h_- = h_u = 0$ and $Dh$ does not appear and so forth.
After some calculation we obtain the following.

\begin{proposition}\label{pr:der3b}
  The most general derivation of the general
  $V_{\text{IIIb}}(E,J,\fl,\fh,\fg,\psi)$ is given by
\begin{equation*}
  \begin{aligned}[m]
    D\be_+ &= \alpha \be_+ + \beta v \\
    D v &= \gamma \be_+ + \delta v \\
    D\be_- &= a \be_- + b u + x_- + h_- + g_- + \eta \be_+ + \xi v \\
    D u &= c \be_- + d u + x_u  + \ell_u + \psi g_- + \theta \be_+ + \omega v\\
    Dx &= \varphi(x) + \half (\alpha + a) x - \left<x_-,x\right> \be_+ - \left<x_u,x\right> v \\
    Dh &= [h_D,h] - a h - \left<h_-,h\right>  \be_+ \\
    D\ell &= [\ell_D,\ell] - c \ell - \left<\ell_u,\ell\right> v \\
    Dg &= [g_D, g] - (a + b \psi) g - \left<g_-,g\right> \be_+ - \left<\psi g_-,g\right> v ~,
  \end{aligned}
\end{equation*}
where $h_D \in \fh$, $\ell_D \in \fl$, $g_D \in \fg$, and $\varphi\in
\fso(E)$ and commutes with $J$.  (In other words, $\varphi \in \fu(E,J)$,
the unitary Lie algebra of orthogonal endomorphisms of $E$ which commute
with $J$.)  In addition, when $E$ is present, we must impose the following
\begin{equation}\label{eq:Epresent}
  a + d = 0 \qquad
  \beta = - c \qquad
  \gamma = - b \qquad
  \alpha + a = \delta + d ~;
\end{equation}
when $\fh$ is present, we impose the following
\begin{equation}\label{eq:Hpresent}
  c = 0 \qquad
  \beta = 0 \qquad
  \alpha = -3a ~;
\end{equation}
when $\fl$ is present, we impose the following
\begin{equation}\label{eq:Lpresent}
  b = 0 \qquad
  \gamma = 0 \qquad
  \delta  = -3c ~;
\end{equation}
and finally when $\fg$ is present, we impose the following
\begin{equation}\label{eq:Gpresent}
  a + b \psi = d + c \psi^{-1}  \qquad
  \alpha + 3a + \left(\gamma + 3b\right) \psi = 0 \qquad
  \beta + \left(\delta + 3a\right) \psi + 3b \psi^2 = 0 ~.
\end{equation}
\end{proposition}

\begin{proof}
  The proof is largely routine, except possibly for one thing.  The
  condition on $\varphi_{EE}$ in \eqref{eq:EndV}, says that
  \begin{equation*}
    \varphi_{EE} \circ J - J \circ \varphi_{EE} = (a + d) J
    \qquad\text{and}\qquad \varphi_{EE} + \varphi_{EE}^t = (\alpha + a)
    \id~.
  \end{equation*}
  Consider now the exponential $M(\tau):= \exp(\tau \varphi_{EE})$ for
  $\tau \in \RR$.  Then the first of the above equations says that
  \begin{equation*}
    M(\tau) J M(\tau)^{-1} = \exp (\tau [\varphi_{EE},-]) J = e^{\tau
      (a+d)} J~.
  \end{equation*}
  Taking determinants, using that $J$ is nondegenerate, we see that
  $e^{2n \tau (a+d)} = 1$, where $2n= \dim E$.  This being true for all
  $\tau \in \RR$ implies that $a+d=0$.  Thus $\varphi_{EE}$ commutes
  with $J$.  We break it into a skewsymmetric part (denoted $\varphi$
  above) and symmetric part, which by the second of the above equations
  on $\varphi_{EE}$ is $\half(\alpha+a)\id$.
\end{proof}

We are interested in those derivations $D$ which preserve the conformal
class of the inner product: $\left<Dv_1,v_2\right> +
\left<v_1,Dv_2\right> = 2 \mu \left<v_1,v_2\right>$.

\begin{proposition}\label{pr:CDer3b}
  The most general conformal derivation of the general
  $V_{\text{IIIb}}(E,J,\fl,\fh,\fg,\psi)$ is given by
  \begin{equation*}
    \begin{aligned}[m]
      D\be_+ &= (2\mu -a) \be_+ - c v \\
      D v &= -b \be_+ + (2\mu - d) v \\
      D\be_- &= a \be_- + b u + x_- + h_- + g_- - \theta v \\
      D u &= c \be_- + d u + x_u  + \ell_u + \psi g_- + \theta \be_+\\
      Dx &= \varphi(x) + \mu x - \left<x_-,x\right> \be_+ - \left<x_u,x\right> v \\
      Dh &= [h_D,h] - a h - \left<h_-,h\right>  \be_+ \\
      D\ell &= [\ell_D,\ell] - c \ell - \left<\ell_u,\ell\right> v \\
      Dg &= [g_D, g] - (a + b \psi) g - \left<g_-,g\right> \be_+ - \left<\psi g_-,g\right> v ~,
    \end{aligned}
  \end{equation*}
  where $h_D \in \fh$, $\ell_D \in \fl$, $g_D \in \fg$, $\varphi\in
  \fu(E,J)$ and where, if $\fg$ is present $\mu = -a - b \psi$ in
  addition to \eqref{eq:Gpresent}, if $\fh$ is present $a = - \mu$ in
  addition to \eqref{eq:Hpresent}, and if $\fl$ is present $c = - \mu$
  in addition to \eqref{eq:Lpresent}.  The skewsymmetric derivations are
  obtained setting $\mu = 0$.
\end{proposition}

\section{Imposing the physical constraints}
\label{sec:no-ghosts}

We will now conclude by revisiting the 3-algebraic criteria set out in
Section~\ref{sec:3-BL} in light of our structural results of
Section~\ref{sec:metric} and our classification of metric Lie 3-algebras
with signature $(2,p)$ in Section~\ref{sec:2p}.  We will select those
$(2,p)$ signature Lie 3-algebras which satisfy the criteria and indicate
how to go about constructing more general metric Lie 3-algebras
satisfying the criteria.  We will focus on three specific criteria:
\begin{itemize}
\item decoupling of negative-norm states, which translates into the
  existence of a maximally isotropic centre;
\item absence of scale, which translates into the existence of
  automorphisms which rescale the inner product; and
\item parity invariance of the lagrangian, which translates into the
  existence of isometric anti-automorphisms.
\end{itemize}

\subsection{Decoupling of negative-norm states}
\label{sec:decoupling-rev}

As discussed in Section \ref{sec:shifts}, the existence of the shift
symmetry used in \cite{BLSNoGhost,GomisSCFT} in order to decouple the
negative-norm states present in the case of metric Lie 3-algebras of
indefinite signature, translates into the existence of a maximally
isotropic centre.  As noted noted in Theorem~\ref{th:2p}, for $(2,p)$
signature only case IIIb admits a maximally isotropic plane in its
centre so that $F_{v_1 ABC} = 0 = F_{v_2 ABC}$ relative to the basis
defined in Section~\ref{sec:brief-BL}.  Case Ia, however, has a
non-vanishing $F_{u_1 u_2 v_2 a}$ component.

The results of Section~\ref{sec:metric} allow us to make a more
general statement.  As stated in Corollary~\ref{co:metric3lie}, every
metric Lie 3-algebra can be constructed out of the one-dimensional and
simple Lie 3-algebras iterating the operations of double extension and
orthogonal direct sum.  It is clear from the structure of a double
extension that double extending by a simple Lie 3-algebra $U$ cannot
result in a maximally isotropic subspace of the centre, for maximally
isotropy means that $U \oplus U^*$ should already contain a maximally
isotropic subspace of the centre, yet for $U$ simple, $U\oplus U^*$
has trivial centre.  This means that any double extension must be by a
one-dimensional algebra.  Since we only double extend by a
one-dimensional algebra, the results of \cite{Lor3Lie} show that any
simple factor of the algebra we are double extending can be factored
out, resulting in a decomposable Lie 3-algebra.  Hence we conclude
that the only indecomposable metric Lie 3-algebras admitting a
maximally isotropic subspace of the centre are the ones constructed
out of the one-dimensional Lie 3-algebra iterating the operations of
orthogonal direct sum and double extension.  This does not mean,
however, that all such algebras have a maximally isotropic centre.
The example of type Ia above shows that it is also necessary to impose
the condition that the Lie algebra structure on the subspace
corresponding to the metric Lie 3-algebra we are double extending,
should also contain a maximally isotropic centre.  Such metric Lie
algebras have been studied in \cite{KathOlbrich2p}.

These remarks give \emph{in principle} a prescription for the
construction of such metric Lie 3-algebras.   We start with a
euclidean abelian Lie 3-algebra $A_1$ and we double extend to $\fD(A_1)
= \RR(u_1,v_1) \oplus A_1$, with nonzero brackets
\begin{equation*}
  [u_1,x_1,y_1] = [x_1,y_1]_1 \qquad\text{and}\qquad
  [x_1,y_1,z_1] = - \left<[x_1,y_1]_1,z_1\right> v_1~,
\end{equation*}
where $[-,-]_1$ defines on $A_1$ a metric Lie algebra structure
$\fw_1$.  The most general such Lie algebra is reductive, whence a
direct sum of semisimple and abelian.  It is, in fact, isomorphic to
$W(\fk\oplus\ft)$, where $\fk$ is compact semisimple and $\ft$ is
abelian.  This will be indecomposable if $\ft=0$, otherwise it is
decomposable.  The most general lorentzian Lie 3-algebra built out of
one-dimensional Lie 3-algebras is therefore isomorphic to $W_2:=
W(\fk) \oplus A_2$ for some compact semisimple Lie algebra $\fk$ and
where $A_2$ is an abelian Lie 3-algebra.  Of course, $\fk=0$, in which
case $W(\fk) \oplus A_2 = A'_2$ is abelian.  This yields all possible
lorentzian Lie 3-algebras without simple factors \cite{Lor3Lie}.

We now consider the double extension of $W_2$ by the one-dimensional
algebra: $\fD(A_2) = \RR(u_2,v_2) \oplus W_2$, with nonzero brackets
\begin{equation*}
  [u_2,x_2,y_2] = [x_2,y_2]_2 \qquad\text{and}\qquad
  [x_2,y_2,z_2] = [x_1,y_2,z_2]_2 - \left<[x_2,y_2]_2,z_2\right> v_2~,
\end{equation*}
where $[-,-]_2$ is a lorentzian Lie algebra structure $\fw_2$ on $W_2$
which leaves invariant the Lie 3-algebra brackets $[-,-,-]_2$ of $W_2$
and which has a maximally isotopic centre.  In particular, $\fw_2$ is
a Lie subalgebra of $\Der^0 W_2$.  One must now determine the possible
Lie subalgebras of $\Der^0 W_2$, as was done in Section~\ref{sec:2p}
except that we only allow those with a maximally isotropic centre.
This yields a list of possible metric Lie 3-algebras with signature
$(2,*)$ and with maximally isotropic centre.  Let $W_3 = \fD(W_2) \oplus
A_3$ be one such algebra.  We must now consider all the possible
double extensions $\fD(W_3)=\RR(u_3,v_3) \oplus W_3$ by a
one-dimensional subalgebra, which has brackets
\begin{equation*}
  [u_3,x_3,y_3] = [x_3,y_3]_3 \qquad\text{and}\qquad
  [x_3,y_3,z_3] = [x_1,y_3,z_3]_3 - \left<[x_3,y_3]_3,z_3\right> v_3~,
\end{equation*}
for some Lie algebra structure $\fw_3$ with brackets $[-,-]_3$ on
$W_3$ which is compatible with the 3-brackets on $W_3$ and with
maximally isotropic centre.  This requires classifying the possible
Lie subalgebras $\fw_3 < \Der^0W_3$, etc...  There is a classification
\cite{KathOlbrich2p} of metric Lie algebras with signature $(2,*)$,
whence it ought to be a matter of patience to classify the metric Lie
3-algebras of signature $(3,*)$.  Going beyond this requires knowing
the metric Lie algebras of signature $(3,*)$ which is still open.
Nevertheless, even if shy of a classification, the above procedure
gives a way of constructing examples.

\subsection{Conformal automorphisms and the coupling constant}
\label{sec:coupling-rev}

As discussed in Section~\ref{sec:coupling}, the absence of scale in the
Bagger--Lambert model is guaranteed by the existence of an automorphism
in the Lie 3-algebra which rescales the inner product.  Infinitesimally,
such automorphisms are generated by derivations $D \in \Der V$ such that
$\left<Dx,y\right> + \left<x,Dy\right> = 2\mu \left<x,y\right>$ for some
$\mu \in \RR^\times$ and for all $x,y\in V$.  Such derivations exist for
a large class of $(2,p)$-signature Lie 3-algebras, as we now show.

Several types of metric Lie 3-algebras in $(2,p)$ signature we have
found admit conformal automorphisms that are completely analogous to the
one noted in Section~\ref{sec:coupling} that was used to fix the
coupling constant in the lorentzian case.  For instance,
Proposition~\ref{pr:AutcVIa} for the type Ia algebras shows that the
appropriate conformal automorphism here would be generated by the
parameter $\beta$, with the same powers as in the lorentzian case.
However, we have determined this class already to be physically
uninteresting on the grounds that it does not obey the shift symmetry
criterion noted above.

As noted at the end of Section \ref{sec:type-III}, there are 9 classes
of indecomposable type IIIb algebras, denoted $V(E,J)$, $V(g,\psi)$ for
$\psi$ not a scalar, $V_{\text{IIIb}}(E,J,\fzero,\fzero,\fg,\psi)$ for
$\psi$ again not a scalar, $V_{\text{IIIb}}(E,J,\fh,\fzero,\fzero,0)$,
$V_{\text{IIIb}}(0,0,\fh,\fzero,\fg,\psi)$,
$V_{\text{IIIb}}(E,J,\fh,\fl,\fzero,0)$,
$V_{\text{IIIb}}(0,0,\fh,\fl,\fg, \psi)$,
$V_{\text{IIIb}}(E,J,\fh,\fzero,\fg, \psi)$, and the general
$V_{\text{IIIb}}(E,J,\fh,\fl,\fg, \psi)$.

It is straight-forward to determine which of these algebras possess
conformal derivations with $\mu \neq 0$, by the use of
Proposition~\ref{pr:CDer3b}.

\begin{proposition}\label{pr:IIIb+cder}
  The following indecomposable type IIIb Lie 3-algebras admit nontrivial
  conformal derivations: $V(E,J)$, $V(\fg,\psi)$ for $\psi$ not a
  scalar, $V_{\text{IIIb}}(E,J,\fh,\fzero,\fzero,0)$ and
  $V_{\text{IIIb}}(0,0,\fh,\fzero,\fg,\psi)$ for any $\psi$.
\end{proposition}

For each case in turn we will now exhibit a conformal derivation and the
automorphism to which it exponentiates.  In all cases, the automorphism
is a simple rescaling of the basis elements.

For $V(E,J)$ we have the following conformal derivation
\begin{equation*}
  \begin{aligned}[m]
    D\be_+ = \mu \be_+ \quad
    D v = 3\mu v \quad
    D\be_- = \mu \be_-\quad
    D u = - \mu u \quad
    Dx = \mu x~,
  \end{aligned}
\end{equation*}
for all $x \in E$.  This clearly exponentiates to the following
conformal automorphism
\begin{equation*}
  (\be_+,v,\be_-,u,x) \mapsto \left(e^\mu \be_+, e^{3\mu} v, e^\mu
    \be_-,e^{-\mu} u, e^\mu x\right)~,
\end{equation*}
for all $x \in E$.

For $V(\fg,\psi)$ and $\psi$ not a scalar, we find the following
conformal derivation
\begin{equation*}
  \begin{aligned}[m]
    D\be_+ = 3\mu \be_+ \quad
    D v = \mu v \quad
    D\be_- = -\mu \be_-\quad
    D u = - \mu u \quad
    D g = \mu g~,
  \end{aligned}
\end{equation*}
for all $g \in \fg$, which exponentiates to the following
conformal automorphism
\begin{equation*}
  (\be_+,v,\be_-,u,g) \mapsto \left(e^{3\mu} \be_+, e^\mu v, e^{-\mu}
    \be_-,e^{-\mu} u, e^\mu g\right)~,
\end{equation*}
for all $g \in \fg$.

For $V_{\text{IIIb}}(E,J,\fh,\fzero,\fzero,0)$, we find the following
conformal derivation
\begin{equation*}
  \begin{aligned}[m]
    D\be_+ = 3\mu \be_+ \quad
    D v = \mu v \quad
    D\be_- = -\mu \be_-\quad
    D u = - \mu u \quad
    D x = \mu x \quad
    D h = \mu h~,
  \end{aligned}
\end{equation*}
for all $x \in E$ and $h \in \fh$, which exponentiates to the following
conformal automorphism
\begin{equation*}
  (\be_+,v,\be_-,u,x,h) \mapsto \left(e^{3\mu} \be_+, e^\mu v, e^{-\mu}
    \be_-,e^{-\mu} u, e^\mu x, e^\mu h\right)~,
\end{equation*}
for all $x \in E$ and $h \in \fh$.

Finally for $V_{\text{IIIb}}(0,0,\fh,\fzero,\fg,\psi)$ and any $\psi$,
we find the following conformal derivation
\begin{equation*}
  \begin{aligned}[m]
    D\be_+ = 3\mu \be_+ \quad
    D v = \mu v \quad
    D\be_- = -\mu \be_-\quad
    D u = - \mu u \quad
    D g = \mu g \quad
    D h = \mu h~,
  \end{aligned}
\end{equation*}
for all $g \in \fg$ and $h \in \fh$, which exponentiates to the following
conformal automorphism
\begin{equation*}
  (\be_+,v,\be_-,u,g,h) \mapsto \left(e^{3\mu} \be_+, e^\mu v, e^{-\mu}
    \be_-,e^{-\mu} u, e^\mu g, e^\mu h\right)~,
\end{equation*}
for all $g \in \fg$ and $h \in \fh$.

\subsection{Parity invariance}
\label{sec:parity-rev}

As discussed in Section~\ref{sec:parity}, parity invariance of the
Bagger--Lambert action demands the existence of an isometric
anti-automorphism of the Lie 3-algebra.  It is easy to find isometric
anti-automorphisms for all four types of indecomposable Lie 3-algebras in
Proposition~\ref{pr:IIIb+cder} admitting nontrivial conformal
automorphisms.  They are given by $\gamma: V \to V$ defined as follows:
\begin{itemize}
\item For $V(E,J)$,
  \begin{equation*}
    \gamma : (\be_+,v,\be_-,u,x) \mapsto  (v, \be_+, u, \be_-, x)~,
  \end{equation*}
  for all $x \in E$;
\item For $V(\fg,\psi)$,
  \begin{equation*}
    \gamma : (\be_+,v,\be_-,u,g) \mapsto  (-\be_+, -v, -\be_-, -u, g)~,
  \end{equation*}
  for all $g \in \fg$;
\item For $V_{\text{IIIb}}(E,J,\fh,\fzero,\fzero,0)$,
  \begin{equation*}
    \gamma : (\be_+,v,\be_-,u,x,h) \mapsto  (\be_+, -v, \be_-, -u, x,-h)~,
  \end{equation*}
  for all $x \in E$ and $h\in\fh$; and
\item $V_{\text{IIIb}}(0,0,\fh,\fzero,\fg,\psi)$,
  \begin{equation*}
    \gamma : (\be_+,v,\be_-,u,h,g) \mapsto  (-\be_+, -v, -\be_-, -u, h,g)~,
  \end{equation*}
  for all $g \in \fg$ and $h\in\fh$.
\end{itemize}

A fuller investigation of the Bagger--Lambert models associated to these
four classes of Lie 3-algebras will be the subject of a forthcoming
preprint.

\bibliographystyle{utphys}
\bibliography{AdS,AdS3,ESYM,Sugra,Geometry,Algebra}

\end{document}